\documentclass[12pt]{article}
\usepackage{graphicx}
\usepackage{amsthm}
\usepackage{amssymb}
\usepackage{amsmath}
\usepackage{amsfonts}
\usepackage{epic}
\usepackage{epsfig}
\usepackage{amscd}

\def\be{\begin{equation}}
\def\ee{\end{equation}}
\input{tcilatex}

\begin{document}

\begin{titlepage}
\begin{flushright}
BRXTH596\\
\end{flushright}
\begin{center}
\noindent{{\LARGE{The Lovelock Black Holes}}}

\smallskip
\smallskip
\smallskip

\smallskip
\smallskip
\smallskip
\smallskip

\smallskip
\smallskip
\noindent{\large{Cecilia Garraffo$^{1,2}$ and Gaston Giribet$^{3,4}$}}

\smallskip
\smallskip

\end{center}
\smallskip
\centerline{$^1$ Brandeis Theory Group, Martin Fisher School of Physics}
\centerline{{\it Brandeis University, Waltham, MA 02454-9110.}}
\smallskip
\smallskip
\smallskip
\centerline{$^2$ Instituto de Astronom\'{\i}a y F\'{\i}sica del Espacio, CONICET}
\centerline{{\it Ciudad Universitaria, C.C. 67 Suc. 28, 1428, Buenos Aires, Argentina.}}
\smallskip
\smallskip
\smallskip
\centerline{$^3$ Department of Physics, Universidad de Buenos Aires and CONICET}
\centerline{{\it Ciudad Universitaria, Pabell\'on I, 1428. Buenos Aires, Argentina.}}
\smallskip
\smallskip
\smallskip
\centerline{$^4$ Centro de Estudios Cient\'{\i}ficos, CECS, Valdivia, Chile}
\centerline{{\it Arturo Prat 514, Valdivia, Chile.}}

\smallskip

\smallskip

\smallskip

\smallskip

\smallskip

\smallskip

\smallskip

\begin{abstract} Lovelock theory is a natural extension of Einstein theory of gravity to higher dimensions, and it is 
of great interest in theoretical physics as it describes a wide class of models. In particular, it describes string 
theory inspired ultraviolet corrections to Einstein-Hilbert action, while admits the Einstein general relativiy and the 
so called Chern-Simons theories of gravity as particular cases. Recently, five-dimensional Lovelock theory has been 
considered in the literature as a working example to illustrate the effects of including higher-curvature terms in the 
context of AdS/CFT correspondence.

Here, we give an introduction to the black hole solutions of Lovelock theory and analyze their most important 
properties. These solutions can be regarded as generalizations of the Boulware-Deser solution of Einstein-Gauss-Bonnet 
gravity, which we discuss in detail here. We briefly discuss some recent progress in understading these and other 
solutions, like topological black holes that represent black branes of the theory, and vacuum thin-shell wormhole-like 
geometries that connect two different asymptotically de-Sitter spaces. We also make some comments on solutions with 
time-like naked singularities. \end{abstract}

\end{titlepage}



\newpage


\section{Introduction}

\subsubsection*{Why higher-curvature corrections?}

It is a common belief that General Relativity, despite its fabulous success
in describing our Universe at middle and large scale, has to be corrected at
short distance. In particular, the apparent tension between Einstein's
theory and quantum field theory supports the idea that General Relativity is
merely an effective model that would be replaced in the UV\ regime by a
different theory, and such a \textit{new} theory would ultimately permit us
to make sense of what we call Quantum Gravity. The natural scale at which
one expects such short distance corrections to manifestly appear is the
Planck scale $l_{P}$, determined by the Newton's coupling constant $%
G=l_{P}^{2}/16\pi $.

At present, the most successful candidate to represent a quantum theory of
gravity is String Theory (or its mother theory, M-theory). In fact, one of
the predictions of string theory is the existence of a massless particle of
spin $2$ whose dynamics at classical level is governed by Einstein equations%
\begin{equation}
R_{\mu \nu }=0.  \label{BASEepsilon}
\end{equation}

In addition, string theory also predicts next-to-leading corrections to (\ref%
{BASEepsilon}), which would be relevant at distances comparable with the
typical length scale of the theory $l_{s}=\sqrt{\alpha ^{\prime }}$. These
short-distance corrections are typically described by supplementing
Einstein-Hilbert action by adding higher-curvature terms \cite{GrossWitten}, correcting General
Relativity in the UV\ regime. As a result, the stringy spin $2$ interaction
turns out to be \textit{finite}, and this raises the hope to finally have
access to a consistent theory of quantum gravity.

To investigate black hole physics in higher-curvature gravity theories, the
first question we have to answer is whether such theories actually induce
short-distance modifications to the black hole geometry or not. Despite
expectations that the inclusion of higher-curvature terms in the
gravitational action yields modifications to General Relativity, it is not
necessarily the case that such modifications manifestly appear in the static
spherically symmetric sector of the space of solutions. In fact, as we will
see below, Schwarzschild geometry usually resists modifications. In turn,
first it is important to identify the theories of gravity that yield
modifications to the spherically symmetric solution.

\subsubsection*{Schwarzschild metric as a persistent solution}

To warm up, let us start by considering a very simple example of
higher-curvature term. Consider the action%
\begin{equation}
S=\frac{1}{16\pi G}\int {d}^{4}x\sqrt{-g}\left( R-2\Lambda +\alpha {R}%
^{2}\right)  \label{Quinta}
\end{equation}%
which corresponds to Einstein-Hilbert action in four dimensions augmented
with the square of the curvature scalar, where $\alpha $ is a coupling
constant with dimensions of $[\alpha ]=$length$^{2}$. This action is a
particular case of the so-called $f(R)$-gravity theories, which are defined
by adding to the Einstein-Hilbert Lagrangian a function of the Ricci scalar $%
f(R)$. It is well known that $f(R)$-gravity theories are equivalent (after
field redefinition that involves a conformal transformation) to General
Relativity coupled to a scalar field $\phi $, provided a suitable
self-interaction potential $V(\phi )$ that depends on the function $f$ (see 
\cite{fR} and references therein). In this sense, these theories are not
different from particular models of quintessence. Here, we are interested 
in
less simple models; however, let us consider (\ref{Quinta}) as the starting
point of our discussion.

A remarkable point is that the theory defined by action (\ref{Quinta})
admits (Anti-) de Sitter-Schwarzschild metric as its static spherically
symmetric solution. In particular, when $\Lambda =0$ the theory still admits
the Schwarzschild solution even for $\alpha \neq 0$, and it is due to the
property $R_{\mu \nu}=0$.

The theory defined by action (\ref{Quinta}) is not the only theory of
gravity that admits Schwarzschild metric as a persistent solution. Actually,
this is a rather common feature of theories with higher-curvature terms. 
In the case of quadratic terms in four dimensions this is an indirect 
consequence of the Gauss-Bonnet theorem\footnote{The simplest 
pure gravitational theory that excludes Schwarzschild solution in four 
dimensions is a cubic contraction of the Weyl tensor \cite{DeserTekin}. 
In dimension $D > 4$, and because the Kretschmann invariant $R_{\mu \nu 
\sigma \delta } R^{\mu \nu \sigma \delta }$ is independent from the 
quadratic scalars $R^2$ and $R_{\mu \nu } R^{\mu \nu }$, quadratic 
deformation of Einstein gravity may exclude Schwarzschild-Tangherlini solution.}. A second
example we can consider is Einstein gravity coupled to conformally invariant
gravity; namely
\begin{equation}
S=\frac{1}{16\pi G}\int {d}^{4}x\sqrt{-{g}}\bigg({R}-2\Lambda +c\ {C}%
_{\alpha \beta \mu \nu }{C}^{\alpha \beta \mu \nu }\bigg),
\end{equation}%
where $c$ is a coupling constant and $C_{\alpha \beta \mu \nu }$ is the Weyl
tensor, whose quadratic contraction reads 
\begin{equation}
C_{\alpha \beta \mu \nu }C^{\alpha \beta \mu \nu }=\frac{1}{3}%
R^{2}-2R_{\alpha \beta }R^{\alpha \beta }+R_{\alpha \beta \mu \nu }R^{\alpha
\beta \mu \nu }.  \label{eq:W^2}
\end{equation}

The equations of motion associated to this action read 
\begin{equation}
{R}_{\mu \nu }-\frac{1}{2}{R}{g}_{\mu \nu }+\Lambda {g}_{\mu \nu }+c{W}_{\mu
\nu }=0,  \label{WEeqs}
\end{equation}%
where ${W}_{\mu \nu }$ is the Bach tensor, 
\begin{eqnarray}
{W}_{\mu \nu } &=&{\square }{R}_{\mu \nu }-\frac{1}{6}{g}_{\mu \nu }{\square 
}{R}-\frac{1}{3}{\nabla }_{\mu }{\nabla }_{\nu }{R}+2{R}_{\mu \rho \nu
\sigma }{R}^{\rho \sigma }  \notag \\
&&-\frac{1}{2}{g}_{\mu \nu }{R}_{\rho \sigma }{R}^{\rho \sigma }-\frac{2}{3}{%
R}{R}_{\mu \nu }+\frac{1}{6}{g}_{\mu \nu }{R}^{2}.  \label{eq:Bach}
\end{eqnarray}

It is easy to show that, when $\Lambda =0$, Scwarzschild metric solves
equations (\ref{WEeqs}) as well. This follows from the fact that Bach tensor
(\ref{eq:Bach}) vanishes if Ricci tensor vanishes, and thus all solutions to
General Relativity are also solutions of (\ref{WEeqs}).

Another example of a modified theory that admits Schwarzschild metric as a
solution is the Jackiw-Pi theory \cite{JP}. This theory has recently
attracted much attention \cite{YunesReview}. It is defined by the action 
\begin{equation}
S=\frac{1}{16\pi G}\int {d}^{4}x\sqrt{-{g}}\left( {R}-2\Lambda +\frac{\theta 
}{4}\,^{\ast }{R}_{\alpha \beta \mu \nu }{R}^{\alpha \beta \mu \nu }\right) ,
\label{eq:JP}
\end{equation}%
where the function $\theta $ is a Lagrange multiplier that couples to the
Pontryagin density $^{\ast }{R}_{\alpha \beta \mu \nu }{R}^{\alpha \beta \mu
\nu }$, constructed via the dual curvature tensor 
\begin{equation*}
^{\ast }{R}_{~\beta }^{\alpha ~\mu \nu }=\frac{1}{2}\frac{{\varepsilon 
}^{\rho
\sigma \mu \nu }}{\sqrt{-g}}{R}_{~\beta \rho \sigma }^{\alpha },
\end{equation*}%
where ${\varepsilon }_{\rho \sigma \mu \nu }$ is the volume 4-form. The
inclusion of the non-dynamical field $\theta $ comes from the fact that the
Pontryagin density $^{\ast }{R}_{\alpha \beta \mu \nu }{R}^{\alpha \beta \mu
\nu }$ is a total derivative. Action (\ref{eq:JP}) is often called
Chern-Simons modified gravity; however, this has to be distinguished from
the Chern-Simons gravitational theories we will discuss in the section 2.

It is not hard to see that the equations of motion derived from action (\ref%
{eq:JP}) are solved by the Schwarzschild metric. Actually, this is because
the Pontryagin density $^{\ast }{R}_{\alpha \beta \rho \sigma }{R}^{\alpha
\beta \rho \sigma }$ of Schwarzschild metric vanishes. In contrast, Kerr
metric has non-vanishing Pontryagin form, and thus it is not a solution of
Jackiw-Pi theory. In fact, the rotating solution of this theory has not yet
been found, and this represents an interesting open problem as the Jackiw-Pi
theory is considered as a phenomenologically viable correction to Einstein
theory.

Summarizing, there are several models that, while
representing short distance corrections to General Relativity, still admit
the Schwarzschild metric as an exact solution. In particular, this implies
that such models can not be the solution to problems like the issue of the
black hole singularity. On the other hand, there are other models which,
still being integrable, do yield deviations from General Relativity
solutions even in the static spherically symmetric sector. In this paper we
will be concerned with one of such models. We will study a very special case
of higher-curvature corrections to Einstein gravity in higher dimensions,
and we will see that substantial modifications to Schwarzschild solution are
found at short distances.

\subsubsection*{Higher-curvature terms in higher dimensions}

In addition to higher-curvature corrections to Einstein theory, string
theory makes other strong predictions about nature. Probably, the most
important ones are the existence of supersymmetry and the existence of extra
dimensions. In fact, one of the requirements for superstring theory to be
consistent is the space-time to have $9+1$ dimensions; and we learn from our
daily experience that six of these extra dimensions have to be hidden
somehow.

This digression convinces us that studying higher-curvature modification of
General Relativity in higher dimensions seems to be important to address the
problem of quantum gravity, at least within the context of string theory.
This is precisely the subject we will study here. More precisely, in this
paper we will investigate how the string inspired higher-curvature
corrections to Einstein-Hilbert action modify the black hole physics in the
UV regime. This turns out to be a very important question since the black
holes are known to be a fruitful arena to explore gravitational 
phenomena beyond the classical level.

The prototypical example we will analyze is $D$-dimensional quadratic
Lovelock Lagrangian. But, first, before introducing this theory, let us
begin by considering a much more general example. Consider the action

\begin{equation}
S =\int {d}^{D}x\sqrt{-g}\left( R-2\Lambda +\alpha {R}%
^{2}+\beta {R}_{\alpha \beta }{R}^{\alpha \beta }+\gamma {R}_{\alpha \beta
\mu \nu }{R}^{\alpha \beta \mu \nu }\right)  \label{eq:Squad}
\end{equation}%
where the constants $\alpha ,$ $\beta ,$ and $\gamma $ are the coupling
constants for each quadratic term. The field equations obtained by varying
the action (\ref{eq:Squad}) with respect to the metric read%
\begin{eqnarray}
0 &=&G_{\mu \nu }+\Lambda {g}_{\mu \nu }+\left( \beta +4\gamma \right)
\square {R}_{\mu \nu }+\frac{1}{2}\left( 4\alpha +\beta \right) g_{\mu \nu
}\square {R+}  \label{EWEM} \\
&&-\left( 2\alpha +\beta +2\gamma \right) \nabla _{\mu }\nabla _{\nu }{R}%
+2\gamma R_{\mu \gamma \alpha \beta }R_{\nu }^{~\gamma \alpha \beta
}+2\left( \beta +2\gamma \right) R_{\mu \alpha \nu \beta }R^{\alpha \beta }+
\notag \\
&&-4\gamma R_{\mu \alpha }R_{\nu }^{~\alpha }+2\alpha RR_{\mu \nu }-\frac{1}{%
2}\left( \alpha {R}^{2}+\beta {R}_{\alpha \beta }{R}^{\alpha \beta }+\gamma {%
R}_{\alpha \beta \gamma \delta }{R}^{\alpha \beta \gamma \delta }\right)
g_{\mu \nu }  \notag
\end{eqnarray}

Action (\ref{eq:Squad}) is the most general quadratic action one can write
down in $D$-dimensions. For $D\leq 4$, the Gauss-Bonnet theorem permits to
fix $\gamma =0$ without loss of generality. In $D>4$, however, three
quadratic invariants are needed to describe the most general Lagrangian of
this type.

For generic values of the coupling constants $\alpha ,$ $\beta $ and $\gamma 
$, the equations of motion (\ref{EWEM}) are fourth-order differential
equations for the metric (i.e. there are terms prportional to $\nabla _{\mu
}\nabla _{\nu }{R}$, $\square {R}$ and $\square {R}_{\mu \nu }$).
Nevertheless, a remarkable property of (\ref{eq:Squad}) is that there exists
one particular choice of the coupling constants $\alpha $, $\beta $ and $%
\gamma $ that results in the cancellation of all these higher order terms,
yielding second order differential equations. It is easy to see that this
choice is $\alpha =\gamma =-\beta /4$, which only gives a non-trivial
modification to Einstein theory for $D>4$. Actually, in $D=5$ and $D=6$ this choice 
corresponds to Lovelock theory (see (\ref{L}) below); namely
\begin{equation}
S_{L} = \int d^{5}x \sqrt{-g}\left( R-2\Lambda +\alpha \left( {R}^{2}-4 
{R}_{\alpha \beta }{R}^{\alpha \beta }+ {R}_{\alpha \beta
\mu \nu }{R}^{\alpha \beta \mu \nu } \right) \right) 
\end{equation}

It is worth emphasizing that this
choice of coupling constants that yields second order equations of 
motion is \textit{unique} (up to a free parameter $\alpha $). This 
feature also holds in $D$ dimensions, and is a consequence of a more 
general result known as \textit{the Lovelock theorem }\cite{Lovelock}.

In this paper we will be mainly concerned with the theory defined by action (%
\ref{eq:Squad}) with $\alpha =\gamma =-\beta /4$. Besides the \textit{%
uniqueness} of the choice $\alpha =\gamma =-\beta /4$, which already makes
this model interesting in its own right, let us say that this is exactly the
effective Lagrangian that appears in the low energy action of heterotic
string theories in the appropriate frame (and in M-theory compactifications
too).

\subsubsection*{Higher-curvature terms from string theory effective action}

Now, let us sketch how the five-dimensional Lovelock theory arises in the
low energy limit of M-theory (and, consequently, of string theory) when the
theory is compactified from $11D$ (resp. $10D$) to $5D$.

M-theory is supposed to be an extension of string theory; a fundamental theory that, in certain regime, 
would flow 
to string theory \cite{MTh}.

This Mother-theory, if it exists, is yet to be found; nevertheless, we do
know what it has to look like in certain low energy limit: it has to look like
eleven-dimensional supergravity augmented with higher-curvature terms. That
is, the bosonic sector of the M-theory effective action is given by the
graviton $g_{\mu \nu }$ (i.e. the metric) and the 3-form gauge field $A_{\mu
\nu \rho }$ (with field strength $F_{\mu \nu \rho \sigma }=\frac{1}{4}%
\partial _{\lbrack \mu }A_{\nu \rho \sigma ]}$). Including the pure
gravitational fourth order corrections $\mathcal{O}$($R^{4}$), this
effective action takes the form\footnote{The eleven-dimensional Newton constant is given by the Planck scale $%
G_{(11D)}=2\pi ^{4}l_{P}^{9}$.} \cite{TSE}
\begin{eqnarray}
S_{\text{M}} &=&\frac{1}{(2\pi )^{5}l_{P}^{9}}\left[ \int d^{11}x\sqrt{g}R%
\text{ }-\frac{1}{48}\int d^{11}x\sqrt{g}F_{\mu _{1}\mu _{2}\mu _{3}\mu
_{4}}F^{\mu _{1}\mu _{2}\mu _{3}\mu _{4}}+\right.  \label{BASEasterisco} \\
&&-\frac{1}{36(4!)^{2}}\int d^{11}x\varepsilon _{\mu _{1}\mu _{2}..\mu
_{11}}A^{\mu _{1}\mu _{2}\mu _{3}}F^{\mu _{4}\mu _{5}\mu _{6}\mu _{7}}F^{\mu
_{8}\mu _{9}\mu _{10}\mu _{11}}+  \notag \\
&&+\frac{l_{P}^{6}}{27}\left( \frac{3}{2^{13}}\int d^{11}x\sqrt{g}t^{\mu
_{1}\mu _{2}...\mu _{8}}t_{\nu _{1}\nu _{2}...\nu _{8}}R_{\quad \mu _{1}\mu
_{2}}^{\nu _{1}\nu _{2}}R_{\quad \mu _{3}\mu _{4}}^{\nu _{3}\nu
_{4}}R_{\quad \mu _{5}\mu _{6}}^{\nu _{5}\nu _{6}}R_{\quad \mu _{7}\mu
_{8}}^{\nu _{7}\nu _{8}}\right.  \notag \\
&&\left. \left. -\frac{1}{2^{16}}\int d^{11}x\sqrt{g}\varepsilon ^{\mu
_{1}\mu _{2}...\mu _{8} \mu_{9} \mu_{10} \mu_{11}}\varepsilon _{\nu _{1}\nu _{2}...\nu _{8}  \mu_{9} \mu_{10} 
\mu_{11}}R_{\quad
\mu _{1}\mu _{2}}^{\nu _{1}\nu _{2}}R_{\quad \mu _{3}\mu _{4}}^{\nu _{3}\nu
_{4}}R_{\quad \mu _{5}\mu _{6}}^{\nu _{5}\nu _{6}}R_{\quad \mu _{7}\mu
_{8}}^{\nu _{7}\nu _{8}}\right) \right] +\text{ }...  \notag
\end{eqnarray}%
where the ellipses stand for the fermionic content and higher-order
contributions. These higher order contributions include terms like $\mathcal{%
O}$($F^{4}$) and also couplings of the form $\mathcal{O}$($A$ $R^{4}$); we
will not consider these terms here. However, let us mention that the existence of the terms ${\mathcal O}(R^4)$ in the 
action above are related to the terms $\mathcal{O}$($A$ $R^{4}$) through supersymmetry, although indirectly.

The tensor $t^{\mu _{1}...\mu _{8}}$ in the third line of (\ref%
{BASEasterisco}) is defined in terms of the way it acts on antisymmetric
tensors of second rank, namely 
\begin{equation*}
t^{\mu _{1}\mu _{2}...\mu _{8}}B_{\mu _{1}\mu _{2}}B_{\mu _{3}\mu
_{4}}B_{\mu _{5}\mu _{6}}B_{\mu _{7}\mu _{8}}=24\text{tr}(B^{4})-6\text{tr}%
(B^{2})^{2},
\end{equation*}%
where tr($B^{n}$) refers to the trace of $B^{n}$.

The term in the fourth line in (\ref{BASEasterisco}) is actually one of the
terms that appear in the Lagrangian of Lovelock theory (see (\ref{L}) below,
where this term is expressed in an alternative way). In contrast, the term
in the third line, which is of the same order, does not correspond to a term
in the Lovelock theory\footnote{Actually, while second-order terms of heterotic string theory expressed in a
particular frame agree with the second-order term of the Lovelock theory,
the fourth-order terms of Type IIA and IIB string theories (and M-theory) do
not agree with the fourth-order term of the Lovelock theory.}.

A string theory ${\mathcal O}(R^4)$ contribution similar to that of the third and fourth lines of (\ref{BASEasterisco}) also
appears in ten dimensions \cite{GrossWitten}. This can be written as follows\footnote{Compactifying to four dimensions gives raise to the 
higher-curvature correction
\begin{equation*}
\int d^{4}x \sqrt{g} \left( \left( ^{\ast }{R}_{\alpha \beta \mu \nu }{R}^{\alpha
\beta \mu \nu }\right) ^{2}+\,\left( {R}_{\alpha \beta \mu \nu }{R}^{\alpha
\beta \mu \nu }\right) ^{2} \right) .
\end{equation*}
See \cite{GruzinovKleban} for a recent discussion on these quartic terms in four dimensions.}
\begin{eqnarray*}
\int d^{10}x \sqrt{g} ( ( R^{\mu \nu \alpha \beta }R_{\mu \nu \alpha
\beta }) ^{2}+2R_{\mu \nu \rho \sigma }R^{\mu \nu \alpha \beta
}R_{\alpha \beta \gamma \delta }R^{\gamma \delta \mu \nu }-8R_{\mu \nu
\alpha \beta }R_{\ \ \gamma \delta }^{\mu \nu }R^{\rho \sigma \beta \gamma
}R_{\ \ \rho \sigma }^{\delta \alpha } \\ 
-16R^{\mu \nu \alpha \gamma }R_{\mu \nu \alpha \beta }R^{\rho \sigma
\delta \beta }R_{\rho \sigma \delta \gamma } +16R^{\rho \nu \gamma \beta
}R_{\mu \nu \alpha \beta }R^{\mu \sigma \alpha \delta }R_{\rho \sigma \gamma
\delta } + 32R^{\rho \nu \gamma \beta }R_{\mu \nu \alpha \beta }R_{\ \ \delta \gamma
}^{\sigma \mu }R_{\ \ \sigma \rho }^{\delta \alpha }  )  .
\end{eqnarray*}

Now, let us analyze what happens when the M-theory effective action we
discussed above (including the higher-curvature terms $\mathcal{O}(R^{4})$)
is compactified to five dimensions. Let us assume we reduce from $11D$ to $%
5D $ by compactifying six of the eleven dimensions in compact Calabi-Yau (CY$_{6}$)
threefold. In that case, the effective action of
the five-dimensional theory takes the form \cite{Minasian, Minasian2}
\begin{equation}
{S}_{\text{eff}}=\int {d}^{5}x\sqrt{-{g}}\bigg({R} + \frac{1}{16}%
c_{(2)}^{I}V_{I}(R^{2}-4R_{\mu \nu }R^{\mu \nu }+{R}_{\alpha \beta \mu \nu }{%
R}^{\alpha \beta \mu \nu })\bigg),  \label{EWEMM}
\end{equation}%
where we used units such that $G_{(5D)}=1/16\pi $, and where the coupling $%
c_{(2)}^{I}V_{I}$ is a quantity that depends on the \textit{details} of the
internal CY$_{6}$ manifold\footnote{
More precisely, $c_{(2)}^{I}$ are the components of the second Chern-class
of the $6D$ Calabi-Yau space, while $V^{I}$ are the so-called scalar
components of the vector multiplet, which are proportional to the K\"{a}hler
moduli of the Calabi-Yau; see also \cite{Strominger}. The quantity  $c_{(2)}^{I}V_{I}$ is given by the integral of 
the 6-dimensional extension of the 4-dimensional Euler characteristic over CY$_6$, namely $c_{(2)}^{I}V_{I} \propto 
\int_{CY_6} d^6y (R^2+R_{\mu \nu \rho \gamma } R^{\mu \nu \rho \gamma } - 4R_{\mu \nu } R^{\mu \nu } ) $. In 
addition, the dimensional reduction of terms ${\mathcal O}(R^4)$ gives raise to other corrections, like the shifting 
of the coefficient of the Einstein-Hilbert term.}.

In turn, we see that quadratic terms in (\ref{EWEMM}) come from the $\mathcal{O}(R^{4})$ terms\footnote{Let us be reminded of the fact that 
M-theory effective action also has other terms of the form $\mathcal{O}(AR^{4})\sim A\wedge ($Tr$R^{4}-\frac{1}{4}($Tr $R^{2})^{2})$.} of 
(\ref{BASEasterisco}). We observe that action (\ref{EWEMM}) resembles a particular case of (\ref{eq:Squad}), namely the case $D=5$ with $\alpha 
=\gamma =-\beta /4$, identifying $c_{(2)}^{I}V_{I}=16\alpha $. This is precisely the theory we will study in this paper: the most general quadratic 
theory of gravity with equations of motion of second order, which, as we have just seen, arises as Calabi-Yau compactifications of M-theory. We already 
mentioned that a quadratic action similar to (\ref{EWEMM}) also appears in the 1-loop corrected effective action of heterotic string theory. Written in 
the Einstein frame, the coupling of higher-curvature terms in the heterotic effective action is given by $\alpha \sim \alpha' e^{\phi}$, where the 
dilaton field $\phi $ clearly contributes. Black holes solutions in dilatonic Einstein-Gauss-Bonnet theory were studied in Refs. \cite{DEGB1, DEGB2, 
DEGB3}.

\subsubsection*{Higher-curvature terms in AdS/CFT correspondence}

Because an action like (\ref{EWEMM}) also appears in the effective action of
the heterotic string, it is also usually referred to as "string inspired
higher-curvature corrections". In turn, it represents a nice model to
explore the effects of next-to-lading contributions of string theory to
gravitational physics. In particular, this five-dimensional (Lovelock) model
of gravity was recently considered in the context of AdS/CFT holographic
correspondence \cite{Malda}. Actually, one of the applications of the Lovelock theory to
AdS/CFT that has attracted attention recently was that of showing that the
so-called Kovtun-Son-Starinets bound \cite{TheKSS,TheKSS2} may be violated in a
theory that contains higher-curvature corrections. The Kovtun-Son-Starinets
bound (KSS) is the conjecture that states: the ratio between the shear
viscosity $\eta $ to the entropy $s$ of all the materials obey the universal
relation%
\begin{equation}
\frac{\eta }{s}\geq \frac{1}{4\pi }  \label{TheKSS}
\end{equation}

In Refs. \cite{MartaK,RHIC,RHIC2} it was observed that when action (\ref%
{eq:Squad}) with $\alpha =\gamma =-\beta /4$ and $\Lambda =-l^{-2}<0$ is
considered in asymptotically locally AdS$_{5}$ space, then the conformal
field theory (CFT) that would be dual to such a theory of gravity would
satisfy%
\begin{equation}
\frac{\eta }{s}=\frac{1}{4\pi }\left( 1-\frac{4\alpha }{3l^{2}}\right)
\end{equation}%
what then would violate (\ref{TheKSS}) for $\alpha >0$. Therefore, the KSS
bound would be violated for all the CFTs with a Einstein-Gauss-Bonnet
gravity duals with positive $\alpha $, and this is precisely the sign of $%
\alpha $ that comes from string theory.

The consideration of five-dimensional Lovelock theory as a working example
to study the effects of including higher-curvature terms in AdS/CFT has been
an active line of research in the last years. Just recently, very
interesting papers discussing the interplay between causality and
higher-curvature terms in the context of AdS/CFT appeared
\cite{Dieguito,MyersApuraste}; Ref. \cite{MyersApuraste} considers the 
case of $D=5$ Lovelock theory. Causality in the dual CFT constrains the value of the coupling of the quadratic
Gauss-Bonnet term\footnote{It is usual to define the dimensionless parameter 
$\lambda=-\Lambda \alpha /3$. In terms of this parameter, the permitted range reads $-7/36 < \lambda < 
9/100 $.}. The causality bound comes from demanding that the {\it recidivist gravitons} that hit back the 
boundary after a bulk excursion do not spoil locality in the CFT. The permitted range for the coupling $\alpha $ turns 
out to be 
\begin{equation}
-\frac{7l^2}{12} < \alpha < \frac{27l^2}{100}  \label{bOUND}
\end{equation}

The value $\alpha=3l^2/4$ (i.e. $\lambda = 1/4$), which is not in this range, corresponds to the
Chern-Simons theory of gravity, which we will discuss in section 2. At this
value, the ration $\eta/s$ would vanish (if it were the case that the theory at
$\lambda = 1/4$ has a dual description too\footnote{We thank D. Hofman and J. Edelstein for conversations about 
the case $\lambda = 1/4$.}).

Other works discussing higher-curvature actions in the context of AdS/CFT appeared recently. See for instance 
\cite{Rut}, where holographic superconductors in five-dimensional Lovelock gravity are considered, showing that 
higher-curvature corrections affect the condensation phenomenon. Besides, the ${\mathcal O}(R^2)$ corrections in the 
non-relativistic version of AdS/CFT \cite{BalasubramanianMcGreevy,DTSon} were also studied, and the $D=5$ Lovelock 
theory is also used in Ref. \cite{AdamsMaloney} as the working example for illustrating the renormalization of the 
dynamical exponent $z$. Let us now move to discuss Lovelock theory in detail.

\subsubsection*{The Lovelock Theory of Gravity}

Lovelock theory is the most general metric theory of gravity yielding
conserved second order equations of motion in arbitrary number of dimensions 
$D$. In turn, it is the natural generalization of Einstein's general
relativity (GR) to higher dimensions \cite{Lovelock, Lovelock2}. In three
and four dimensions Lovelock theory coincides with Einstein theory \cite%
{Lanczos}, but in higher dimensions both theories are actually different. In
fact, for $D>4$ Einstein gravity can be thought of as a {\it particular} case of
Lovelock gravity since the Einstein-Hilbert term is one of several terms
that constitute the Lovelock action. Besides, Lovelock theory also admits
other quoted models as particular cases; for instance, this is the case of
the so called Chern-Simons gravity theories, which in a sense are actual
gauge theories of gravity.

On the other hand, Lovelock theory resembles also string inspired models of
gravity as its action contains, among others, the quadratic Gauss-Bonnet
term, which is the dimensionally extended version of the four-dimensional
Euler density. This quadratic term is present in the low energy effective
action of heterotic string theory \cite{CallanKlebanovPerry,
CandelasHorowitzStromingerWitten, GrossSloan}, and it also appears in
six-dimensional Calabi-Yau compactifications of M-theory; see \cite%
{Strominger} and references therein. In \cite{Zwiebach} Zwiebach earlier
discussed the quadratic Gauss-Bonnet term within the context of string
theory, with particular attention on its property of being free of ghost
about the Minkowski space. Besides, the theory is known to be free of ghosts
about other exact backgrounds \cite{BoulwareDeser}. For a nice and concise
review on stringy corrections to gravity actions \cite{CallanMyersPerry,
Myers, Myers2}\ see the introduction of \cite{Schiappa} and references
therein. For interesting recent discussions on higher order curvature terms
see \cite{Strominger,GruzinovKleban,Otros, Fachoignorante,otros2} and
related works.

The Lovelock theory represents a very interesting scenario to study how the
physics of gravity results corrected at short distance due to the presence
of higher order curvature terms in the action. In this paper we will be
concerned with the black hole solutions of this theory, and we will discuss
how short distance corrections to black hole physics substantially change
the qualitative features we know from our experience with black holes in GR.
So, let us introduce the Lovelock theory.

The Lagrangian of the theory is given as a sum of dimensionally extended
Euler densities, and it can be written as follows\footnote{%
Here we are ignoring the boundary terms. We will consider these terms in
section 2.} \cite{Lovelock, Lovelock2}%
\begin{equation}
\mathcal{L}=\sqrt{-g}\ \sum\limits_{n=0}^{t}\alpha _{n}\ \mathcal{R}%
^{n},\qquad \mathcal{R}^{n}=\frac{1}{2^{n}}\delta _{\alpha _{1}\beta
_{1}...\alpha _{n}\beta _{n}}^{\mu _{1}\nu _{1}...\mu _{n}\nu
_{n}}\prod\limits_{r=1}^{n}R_{\quad \mu _{r}\nu _{r}}^{\alpha _{r}\beta _{r}}
\label{L}
\end{equation}%
where the generalized Kronecker $\delta $-function is defined as the
antisymmetric product 
\begin{equation}
\delta _{\alpha _{1}\beta _{1}...\alpha _{n}\beta _{n}}^{\mu _{1}\nu
_{1}...\mu _{n}\nu _{n}}=\frac{1}{n!}\delta _{\lbrack \alpha _{1}}^{\mu
_{1}}\delta _{\beta _{1}}^{\nu _{1}}...\delta _{\alpha _{n}}^{\mu
_{n}}\delta _{\beta _{n}]}^{\nu _{n}}.
\end{equation}

Each term $\mathcal{R}^{n}$ in (\ref{L}) corresponds to the dimensional
extension of the Euler density in $2n$ dimensions\footnote{%
The $2n$-dimensional Euler density $\chi $ is given by $\chi ($M$)=\frac{%
(-)^{n+1}\Gamma (2n+1)}{2^{2+n}\pi ^{n}\Gamma (n+1)}\int_{\text{M}}d^{2n}x%
\sqrt{-g}\ \mathcal{R}^{n}$, where, again, we are not considering the
boundary terms.}, so that these only contribute to the equations of motion
for $n<D/2$. Consequently, without lack of generality, $t$ in (\ref{L}) can
be taken to be $D=2t$ for even dimensions and $D=2t+1$ for odd dimensions%
\footnote{%
See \cite{QWERT5} for a related discussion on gravitational dynamics and
Lovelock theory.}.

The coupling constants $\alpha _{n}$ in (\ref{L})\ have dimensions of
[length]$^{2n-D}$, although it is convenient to normalize the Lagrangian
density in units of the Planck scale $\alpha _{1}=(16\pi G)^{-1}=l_{P}^{2-D}$%
. Expanding the product in (\ref{L}) the Lagrangian takes the familiar form%
\begin{equation}
\mathcal{L}=\sqrt{-g}\ (\alpha _{0}+\alpha _{1}R+\alpha _{2}\left(
R^{2}+R_{\alpha \beta \mu \nu }R^{\alpha \beta \mu \nu }-4R_{\mu \nu }R^{\mu
\nu }\right) +\alpha _{3}\mathcal{O}(R^{3})),  \label{L2}
\end{equation}%
where we see that coupling $\alpha _{0}$ corresponds to the cosmological
constant $\Lambda $, while $\alpha _{n}$ with $n\geq 2$ are coupling
constants of additional terms that represent ultraviolet corrections to
Einstein theory, involving higher order contractions of the Riemann tensor $%
R_{\quad \mu \nu }^{\alpha \beta }$. In particular, the second order term $%
\mathcal{R}^{2}=R^{2}+R_{\alpha \beta \mu \nu }R^{\alpha \beta \mu \nu
}-4R_{\mu \nu }R^{\mu \nu }$ is precisely the Gauss-Bonnet term discussed
above. The cubic term\footnote{cf. \cite{TseytlinSeVolvioCubico}, where it was shown that no unambiguous cubic terms arise in string theory effective 
action; in particular, the Lovelock cubic term is studied. Cubic terms are strongly constrained by supersymmetry.} 
still has a moderate form \cite{MullerHoissen}, namely
\begin{eqnarray}
\mathcal{R}^{3} &=&R^{3}+3RR^{\mu \nu \alpha \beta }R_{\alpha \beta \mu \nu
}-12RR^{\mu \nu }R_{\mu \nu }+24R^{\mu \nu \alpha \beta }R_{\alpha \mu
}R_{\beta \nu }+16R^{\mu \nu }R_{\nu \alpha }R_{\mu }^{\alpha }+  \notag \\
&&+24R^{\mu \nu \alpha \beta }R_{\alpha \beta \nu \rho }R_{\mu }^{\rho
}+8R_{\ \ \alpha \rho }^{\mu \nu }R_{\ \ \nu \sigma }^{\alpha \beta }R_{\ \
\mu \beta }^{\rho \sigma }+2R_{\alpha \beta \rho \sigma }R^{\mu \nu \alpha
\beta }R_{\ \ \mu \nu }^{\rho \sigma }.  \label{cubo}
\end{eqnarray}%
The fourth order term $\mathcal{R}^{4}$ coincides with the pure gravitational term in the last line of (\ref{BASEasterisco}).

Even though the way of writing Lovelock action in its tensorial form (\ref%
{L2})-(\ref{cubo}) may result clear to introduce the theory, it is not the
most efficient way for most of the calculations one usually deal with. A
more convenient way of working out these expressions is to resort to the
so-called first-order formalism, which turns out to be useful both for
formal purposes and for practical ones. Nevertheless, it is important to
point out that the first-order formalism is not necessarily equivalent to
the second-order formalism, so it should not be regarded merely as a
different nomenclature. In the first-order formalism, both the vielbein $%
e_{\mu }^{a}$ and the spin connection $\omega _{\mu }^{ab}$ are considered
as independent degrees of freedom, and the torsion acquires in general
propagating degrees of freedom \cite{T9907109}. It is only in the
torsion-free sector where both formulations are equivalent; notice that the
vanishing torsion condition is always allowed by the equations of motion;
see \cite{Zanelli}, see also \cite{QWERT1}. We will make use of the
first-order formalism at the end of section 2, as it is almost unavoidable
in the discussion of Chern-Simons theory. However, with the intention to
make the exposition as friendly as possible, we will avoid abstruse notation
in the rest of the paper. In any case, since we could not afford to give all
the definitions necessary to introduce the subject, we will assume the
reader is familiarized with basic notions of the theory of gravity and with
the standard nomenclature.

\subsubsection*{Overview}

The paper is organized as follows. In section 2, we analyze the spherically
symmetric black hole solutions in Lovelock theory \cite{Wheeler,Wheeler2}.
In five-dimensions this is given by the Boulware-Deser solution \cite%
{BoulwareDeser}, whose most important properties we review. The special
properties of electrically charged black holes \cite{Wiltshire,Wiltshire2}
are also briefly discussed. In one of the subsections of section 2, we
extend the analysis to those black objects whose horizon geometries
correspond to more general spaces of constant (but not necessarily positive)
curvature \cite{T0011097,R41}. These are the so-called topological black
holes, which can be thought of as black brane solutions of the theory. Also,
we briefly review the most relevant features of the Lovelock black hole
thermodynamics \cite{R46}, focusing our attention on the qualitative
features that have no analogue in GR. Throughout the discussion, the
five-dimensional black hole of the Einstein-Gauss-Bonnet theory will serve
as prototypical example. In section 3, we discuss the role of boundary terms 
\cite{MyersBoundary} and the junction conditions these yield \cite{Davis,
GravanisWillison, GravanisWillison2}. We show how solutions with non-trivial
topology can be constructed by a method of a geometric surgery. Particular
attention is focussed on vacuum wormhole solutions recently found \cite%
{GravanisWillison3, GGGW}. Finally, we study the spherically symmetric
solutions that develop naked curvature singularities. We study these naked
singularities with quantum probes and show that, in spite of the divergence
in the curvature, these spaces are well-behaved within a quantum mechanical
context.

\section{The Lovelock black holes}

\subsubsection*{Spherically symmetric black hole solutions}

Let us first consider the theory in five dimensions. Since in $D<7$ the $%
\mathcal{R}^{3}$ term does not contribute to the equations of motion, the
five-dimensional Lovelock theory basically corresponds to Einstein gravity
coupled to the dimensional extension of the four dimensional Euler density,
i.e. the theory that is usually referred as Einstein-Gauss-Bonnet theory
(EGB). The spherically symmetric static solution of EGB theory was obtained
by Boulware and Deser in Ref. \cite{BoulwareDeser}. The metric takes the
simple form \cite{QWERT2}%
\begin{equation}
ds^{2}=-V^{2}(r)dt^{2}+V^{-2}(r)dr^{2}+r^{2}d\Omega _{3}^{2}
\label{intervalo}
\end{equation}%
where $d\Omega _{3}^{2}$ is the metric of a unitary $3$-sphere, and where
the metric function $V^{2}(r)$ is given by%
\begin{equation}
V^{2}(r)=1+\frac{r^{2}}{4\alpha }+\sigma \frac{r^{2}}{4\alpha }\sqrt{1+\frac{%
16\alpha M}{r^{4}}+\frac{4\alpha \Lambda }{3}},  \label{sol}
\end{equation}%
with $\sigma ^{2}=1$. Here we used the standard convention $\alpha
_{0}/\alpha _{1}=-2\Lambda $, $\alpha _{2}/\alpha _{1}=\alpha $, and,
besides, we have set the Newton constant to a specific value for short. From
(\ref{sol}) we notice that there exist two different branches of solutions
to the spherically symmetric ansatz (\ref{intervalo}), namely $\sigma =+1$
and $\sigma =-1$, and this reflects the fact that the equations of motion
give a differential equation quadratic in the metric function\ $V^{2}(r)$.
As usual, the parameter $M$ arises here as an integration constant, and it
corresponds to the mass of the solution\footnote{%
For the discussion on the computation of charges in this theory see the list
of references \cite%
{R45,T0601081,T0412046,Tcargas,Tcargas2,T0405267,R30,R21,R9,DeserMasa}; see
also \cite{H0303082,H0501044,H0212292,H0310098,H08011021}.}, up to the
factor we absorbed\footnote{%
More precisely, in the definition of $M$ we absorbed a factor $\frac{8\pi G}{%
(D-2)\Omega _{D-2}}$ where $\Omega _{n}=\frac{(n+1)\pi ^{(n+1)/2}}{\Gamma
((n+3)/2)}$ is the surface of the $n$-sphere, and where $G$ is the Newton
constant, given by $G\sim \alpha _{1}^{-1}$, which has been fixed to a
specific values such that $\alpha _{1}=1$.} in $M$.

It is worth mentioning that (\ref{intervalo})-(\ref{sol}) is the most
general spherically symmetric solution to EGB theory, provided the fact that
the metric is smooth everywhere and that the parameters $\Lambda $ and $%
\alpha $ are generic enough. In turn, a Birkhoff theorem holds for this
model \cite{Deser, Zegres, Cristo, DeserTekin}. It is important to emphasize
that for very particular choices of the set of parameters $\alpha _{n}$,
degeneracy in the space of solutions can appear, and in those special cases
the Birkhoff's theorem can be circumvented; see \cite{Zegres} for a very
interesting discussion. To our knowledge, the most complete analysis of the
EGB analogue of Birkhoff's theorem was performed in \cite{R9}, where the
Nariai-type solutions \cite{H0401192} where also discussed.

If $\alpha >0$, the solution corresponding to $\sigma =-1$ in (\ref{sol})
may represent a black hole solution whose horizon, in the case $\Lambda =0$,
is located at $r_{+}=\sqrt{2(M-\alpha )}$. On the other hand, as long as $%
\alpha >0$ and $M>0$, the branch $\sigma =+1$ has no horizon but presents a
naked singularity at $r=0$.

Solutions $\sigma =-1$ and $\sigma =+1$ have substantially different
behaviors, and only one of them tends to the GR solution in the small $%
\alpha $ limit. In fact, in the limit $\alpha \rightarrow 0$ the branch $%
\sigma =-1$ looks like 
\begin{equation}
V_{\sigma =-1}^{2}(r)\simeq 1-\frac{2M}{r^{2}}-\frac{\Lambda }{6}r^{2},
\label{sol1}
\end{equation}%
where we see it approaches the five-dimensional (Anti)-de
Sitter-Schwarzschild-Tangherlini solution \cite{Tangherlini}. On the other
hand, in the $\alpha \rightarrow 0$ limit the solution corresponding to the
branch $\sigma =+1$ behaves like 
\begin{equation}
V_{\sigma =+1}^{2}(r)\simeq 1+\frac{2M}{r^{2}}+\frac{\Lambda }{6}r^{2}+\frac{%
1}{2\alpha }r^{2},  \label{sol5}
\end{equation}%
and we see it acquires a large effective cosmological constant term $\sim
r^{2}/2\alpha $. In particular, this implies that microscopic (A)dS
space-time is a solution of the theory even for $\Lambda =0$. This feature
was expressed by Boulware and Deser \cite{BoulwareDeser}\ by saying that EGB
theory has its own cosmological constant problem, with $\Lambda _{\text{eff}%
}\sim -1/\alpha $. In a sense, the branch $\sigma =+1$ is commonly believed
to be a false vacuum of the theory, and it is known to present ghost
instabilities \cite{BoulwareDeser}; see also \cite{BoulwareDeser2}.

The branch $\sigma =-1$, on the other hand, is well-behaved, and it
represents short distance corrections to GR black holes (\ref{sol1}). While
at short distances the black hole solutions of both theories are
substantially different due to the effects of the Gauss-Bonnet term, in the
large distance regime $r^{2}>>\alpha $ the Lovelock black hole (\ref%
{intervalo}) with $\sigma =-1$ behaves like a GR black hole whose parameters 
$M$ and $\Lambda $ get corrected by finite-$\alpha $ subleading
contributions $\mathcal{O}(\alpha \Lambda )$; for instance, the parameter of
the mass term gets corrected yielding the effective mass $M\sqrt{1+4\alpha
\Lambda /3}$. In the large $r$ limit, the next-to-leading $r$-dependent
contribution to (\ref{sol1}) goes like $\mathcal{O}(\alpha r^{-6})$. 

The damping of this additional term, which in $D$ dimensions goes like $\mathcal{%
O}(\alpha r^{4-2D})$, is actually strong, and, for distance large enough, it
is negligible even in comparison with semiclassical corrections to the
metric due to field theory backreaction, which typically go like $\mathcal{O}%
(\hbar r^{5-2D})$ (for instance, see \cite{Alan}).

All these features are essentially due to the nature of the Gauss-Bonnet
term, and also hold in higher dimensions. In fact, it is straightforward to
generalize solution (\ref{intervalo}) to the case of EGB gravity in $D>5$
dimensions, and the metric is seen to adopt a very similar form \cite{BoulwareDeser}. Actually, it is given by simply replacing the element of
the $3$-sphere in (\ref{intervalo}) by the element of the unitary $(D-2)$%
-sphere $d\Omega _{D-2}^{2}$, and by replacing the piece $16M\alpha /r^{4}$
in (\ref{sol}) by $16M\alpha /r^{D-1}$.

In spite of the non-polynomial form of (\ref{sol}), the horizon structure of
Boulware-Deser solution is quite simple, and in $D$ dimensions the horizon
location is given by the roots of the polynomial 
\begin{equation}
\frac{\Lambda }{6}r^{D-1}-r^{D-3}-2\alpha r^{D-5}+2M=0, \label{FromEsta}
\end{equation}
where $\Lambda $ has been appropriately rescaled by a \thinspace $D$%
-dimensional constant factor.

From (\ref{FromEsta}) we observe that the five-dimensional case is actually a remarkable example since, among
other special features, it allows to have massive solutions with naked
singularities. We mentioned above that if $D=5$ and $\Lambda =0$ the black
hole horizon is located at $r_{+}^{2}=2(M-\alpha )$, and this implies a
lower bound for the spherical solution not to develop a naked singularity,
namely $M>\alpha $. That is, for $0<M<\alpha $ we do find naked
singularities even for the well-behaved branch $\sigma =-1$ with positive $M$%
. For the model with a second order term $\mathcal{R}^{2}$ this only occurs
in $D=5$. In seven dimensions, for instance, the Boulware-Deser solution
with $\Lambda =0$ develops horizons at $r_{+}^{2}=\alpha \sqrt{1+2M/\alpha
^{2}}-\alpha $ and then the horizon always exists provided $\alpha >0$, $M>0$%
. Naked singularities in $D=2n+1$ dimensions usually arise when a term of
order $\mathcal{R}^{n}$ is present in the action. So, for the EGB theory
this only occurs for $D=5$.

Another special feature of the (uncharged) five-dimensional case is that the
metric (\ref{intervalo}) turns out to be finite at the origin, namely $%
V_{(r=0)}^{2}=1+\sigma \sqrt{M/\alpha }$. Nevertheless, the curvature still
diverges at the origin, although not in a dramatic way. We will return to
this point in the last section where we will discuss naked singularities.

It could be important to mention that the analysis of the dynamical
stability of EGB black holes is also special for $D=5$. The stability
analysis under tensor mode perturbations has been explored recently, and it
has been shown that the EGB\ theory exhibits some differences with respect
to Einstein theory; at least, it seems to be the case for sufficiently small
values of mass in five and six dimensions \cite{R1} where instabilities
arise; see also Refs. \cite{R2, DottiGleiser3,R4,R18}. In this sense, the
cases $D=5$ and $D=6$ are special ones. See Ref. \cite{H08043694} for an
interesting recent discussion. On the other hand, let us be reminded of the fact that in $D>6$ dimensions the Lovelock
action (\ref{L}) presents also additional terms of higher order $n>2$, so
that in $D\geq 7$ the Boulware-Deser black hole geometry (\ref{intervalo})-(%
\ref{sol}) only corresponds to a very special example of Lovelock black hole.

Spherically symmetric solutions in higher dimensions containing an arbitrary
higher order terms $\mathcal{R}^{n}$ in (\ref{L}) can be implicitly found by
solving a polynomial equation of degree $n$ whose solutions give the metric
function $V^{2}(r)$; this was originally noticed by Wheeler in \cite%
{Wheeler,Wheeler2}. Moreover, several explicit examples containing arbitrary
amount of terms $\mathcal{R},\mathcal{R}^{2},...$ $\mathcal{R}^{n-1},%
\mathcal{R}^{n}$ are also known. These correspond to particular choices of
the couplings $\alpha _{n}$ in (\ref{L}). One of these explicitly solvable
cases corresponds to the Chern-Simons theory, which exists in odd
dimensions. We will briefly discuss this special case below. A remarkable
fact is that in the case a term $\mathcal{R}^{n}$ of the Lovelock expansion (%
\ref{L}) is considered in the action, then the spherically symmetric
solution may still take a very simple expression, and, depending on the
coupling constants $\alpha _{n}$, it may merely correspond to replacing the
square root in (\ref{sol}) by a power $1/n$; see \cite{R34,R28,R53,CTZ} for
explicit examples.

\subsubsection*{Adding electric charge}

On the other hand, it is quite remarkable that electrically charged black
hole solutions in Lovelock theory also present a very simple form. The
solutions charged under both Maxwell and Born-Infeld electrodynamics have
been known for long time \cite{Wiltshire,Wiltshire2}, and these solutions
were reconsidered recently \cite{AFG}. In general, the metric function of a
charged solution takes the form (\ref{sol}) but replacing the mass parameter 
$M$ by a mass function $M(r)$ that depends on the radial coordinate $r$.
Function $M(r)$ depends on the particular electromagnetic Lagrangian one
considers. In the case of Maxwell theory, and in five dimensions, this
function is given by the energy contribution $M(r)\sim \int_{\varepsilon
}^{r}dr\ Q^{2}/r^{3}\sim -Q^{2}/r^{2}+M_{0}$, where $Q$ represents the
electric charge of the black hole, and where the UV cut-off in the integral
is absorbed in the definition of the additive constant $M_{0}$. More
precisely, for charged black holes in Einstein-Gauss-Bonnet-Maxwell theory
we have $M(r)-M_{0}=-Q^{2}/6r^{2}$, as it was originally noticed by
Wiltshire \cite{Wiltshire}. On the other hand, in the case of black holes
charged under Born-Infeld theory, the function $M(r)$ is given by%
\begin{equation}
M(r)-M_{0}=\frac{2}{3}\beta ^{2}\int_{0}^{r}ds\sqrt{s^{6}+\beta ^{-2}Q^{2}}-%
\frac{1}{6}\beta ^{2}r^{4},  \label{SA}
\end{equation}%
where the $\beta ^{2}$ is the Born-Infeld parameter, according to the
standard form of the Lagrangian $\mathcal{L}_{BI}=\beta ^{2}-\beta ^{2}\sqrt{%
1+F^{2}/\beta ^{2}}$. \ In the large $\beta $ limit $\mathcal{L}_{BI}\simeq -%
\frac{1}{2}F^{2}+\mathcal{O}(F^{4}/\beta ^{2})$, and then the metric
approaches the charged solution for the Maxwell-Einstein-Gauss-Bonnet
theory, 
\begin{equation}
M(r)-M_{0}\simeq -\frac{Q^{2}}{6r^{2}}+\mathcal{O}(Q^{4}/r^{8}\beta ^{2}).
\end{equation}

As expected, the five-dimensional Reissner-Nordstr\"{o}m black hole is
recovered in the large $r$ regime for the case $\sigma =-1$.

Charged solutions of Lovelock theory coupled to Born-Infled electrodynamics
present curious features that are not present in the case of
Einstein-Maxwell theory. Perhaps the most relevant one is the existence of
single-horizon charged solutions \cite{AFG}. Besides, Lovelock black holes
charged under Maxwell electrodynamics, and for certain values of the
coupling constants $\alpha _{n}$, can develop curvature singularities at
fixed values of the radial coordinate \cite{CTZ}, making necessary to
exclude a region of the space. This kind of divergence is usually called
branch singularity, and it can also be present in uncharged solutions, as it
happens for solutions of EGB gravity with $M<0$ and $\alpha >0$, \cite%
{R12,R13}.

As in the case of Hoffmann's solution in Born-Infeld-Einstein \cite{Hoffmann} theory, the Lovelock black holes charged 
under Born-Infled theory induce a
contribution to the mass coming from the finite concentration of
electromagnetic energy around the singularity. Of course, this happens
because both theories coincides at large distances. For finite values of $%
\beta $, $M(r)$ has a large distance behavior that induces a mass
contribution $\Delta M=(2\beta ^{2}/3)\int_{0}^{\infty }dr\sqrt{%
r^{6}+Q^{2}\beta ^{-2}}$. In particular, this implies that, for certain
range of $\beta $ and $Q$, naked singularities in five dimensions may arise
even for values of the effective mass $M_{0}+\Delta M$ grater than $\alpha $%
. Notice that the cosmological constant term also acquires a $\beta $%
-dependent contribution $\sim \beta ^{2}$.

In the next subsection we will consider a generalization of the black hole
solutions reviewed here. We will discuss extended black objects in EGB
theory.

\subsubsection*{Topological black holes}

One of the interesting aspects of Lovelock theory is that it admits another
class of black objects, whose horizons are not necessarily positive
curvature hypersurfaces \cite{R41}. These solutions are usually called
topological black holes, and their metric are obtained by replacing the $%
(D-2)$-sphere $d\Omega _{D-2}^{2}$ in (\ref{intervalo}) by a base manifold $%
d\Sigma _{D-2}^{2}$ of constant (but not necessarily positive) curvature,
provided a suitable shifting in the metric function $V^{2}(r)$. Namely,
these solutions read\footnote{These are analogues of the topological black holes
previously known in four-dimensions, which, at constant $t$ hypersurfaces, correspond to fibrations of (closed) base
manifolds $\Sigma _2/\Gamma$ with non-trivial topology.}
\begin{equation}
ds^{2}=-K^{2}(r)dt^{2}+K^{-2}(r)dr^{2}+r^{2}d\Sigma _{D-2}^{2}  \label{Siete}
\end{equation}%
where the metric function is now given by $K^{2}(r)=V^{2}(r)+k-1,$ with $k=-1,0,+1$, being its sign that of 
the curvature of 
the horizon hypersurface, whose line element is $r_{+}^{2}d\Sigma _{D-2}^{2}$. For $k=+1$ the Boulware-Deser solution 
(\ref%
{intervalo})-(\ref{sol}) is recovered. In general, the base manifold $%
d\Sigma _{D-2}^{2}$ here may be given by a more general constant curvature
space: For instance, it can be given by the product of hyperbolic spaces $%
d\Sigma _{D-2}^{2}=dH_{D-2}^{2}$ for the case of negative curvature $k=-1$,
or merely by a flat space piece $d\Sigma _{D-2}^{2}=dx_{i}dx^{i}$. In turn,
solutions (\ref{Siete}) correspond to black brane type geometries. Such
black objects represent fibrations over constant curvature $\left(
D-2\right) $-dimensional hypersurfaces, implying that the event horizon, in
the cases it exists, is not necessarily a compact simply connected manifold.

Consider for example the five-dimensional EGB theory with negative
cosmological constant $\Lambda <0$, and its black brane solution of the form%
\begin{equation}
ds^{2}=-K_{(k=0)}^{2}(r)dt^{2}+K_{(k=0)}^{-2}(r)dr^{2}+r^{2}dx^{i}dx_{i}
\label{Brane}
\end{equation}%
with 
\begin{equation}
K_{(k=0)}^{2}(r)=\frac{r^{2}}{4\alpha }-\sqrt{\frac{r^{4}}{16\alpha ^{2}}%
\left( 1-4|\Lambda |\alpha /3\right) +\frac{M}{\alpha }},
\end{equation}%
where $x^{i}=x^{1},x^{2},x^{3}$. These objects (brane-like configurations
and topological black holes) have attracted some attention recently due to
their curious properties, and, more recently, these were considered in
applications to AdS/CFT; see for instance \cite{RHIC, RHIC2}.

In \cite{R6}, an exhaustive classification of static topological black hole
solutions of five-dimensional Lovelock theory was presented. The authors
considered an ansatz such that spacelike sections are given by warped
product of the radial coordinate $r$ and an arbitrary base manifold $d\Sigma
_{D-2}^{2}$, and they showed that, for values of the coupling constant $%
\alpha _{2}$ generic enough, the base manifold must be necessarily of
constant curvature, and then the solutions of the theory reduce to the
topological extension of the Boulware-Deser metric of the form (\ref{Siete}%
). In addition, they showed that for the special case where the coupling $%
\alpha _{2}$ is appropriately tuned in terms of the cosmological constant $%
\alpha _{0}$, then the base manifold could admit a wider class of
geometries, and such enhancement of the freedom in choosing $d\Sigma
_{D-2}^{2}$ allows to construct very curious solutions with non-trivial
topology. We will return to this point in section 2.

The existence of black holes with generic horizon structure was also
analyzed in \cite{R3}, where selection criteria for the base manifold $%
d\Sigma _{D-2}^{2}$ were discussed\footnote{The authors of \cite{R3}
derived a necessary constraint to be
obeyed by the Euclidean manifold that is candidate to represent a horizon
geometry of a black hole solution in $D$-dimensional Einstein-Gauss-Bonnet
theory. They proved that such a $D-2$-manifold has to obey the equation
$C_{ki}^{\ lm} C_{lm}^{\ kj} \propto \delta ^j_i$, where $C_{i\ kl}^{\ 
j}$ is the Weyl tensor in $D-2$ dimensions.}, and the authors concluded 
that sensible
physical models strongly restrict most of the examples of exotic black holes
with non-constant curvature horizons. Moreover, the different horizon
structures were also studied in \cite{T0011097,R13} together with its
relation to the asymptotic behavior of the corresponding solutions; see also 
\cite{R36,R35,R31,R19,R8}. Recently, the electrically charged topological
black hole solutions were also analyzed, both for the case of the second
order Lovelock theory in \cite{R12,R36} and for the case of the third order%
\footnote{%
Recently, references \cite{R54,R33,R32,Nueva,Nueva2,QWERT8} discussed other
classes of solutions. We will not comment on these solutions here.} Lovelock
theory in \cite{R19}.


One of the most interesting aspects of these objects with non-trivial
horizon geometries is that they enable us to construct a very simple class of
Kaluza-Klein black holes with interesting properties from the
four-dimensional viewpoint. For instance, such a solution was recently
studied by Maeda and Dadhich in Ref. \cite{R10}. These Kaluza-Klein black
holes are given by a product M$_{4}\times $H$_{D-4}$ between a
four-dimensional manifold M$_{4}$ and a $(D-4)$-dimensional hyperbolic space
H$_{D-4}$. It turns out that the four-dimensional piece of the geometry
asymptotically approaches the charged black hole in locally AdS$_{4}$ space.
In turn, the Gauss-Bonnet term acts by emulating the Reissner-Nordstr\"{o}m
term for large $r$, while it changes the geometry at short distances \cite%
{R15,R16,R17}. In addition to these solutions, other exotic Kaluza-Klein
Lovelock black hole solutions with arbitrary order terms of the form $%
\mathcal{R}^{n}$ and for a specific values of the coefficients $\alpha _{n}$%
\ were studied in \cite{R24}. These black holes are different from those
studied in \cite{R10}, and are obtained by considering black\ $p$-brane
geometries of the form M$_{D-p}\times $T$^{p}$ in the Lovelock theory with $%
\alpha _{i}=\delta _{i,n}$ and $2n=D-p$. These solutions exist for $D-p$
even, and, in addition, the horizon structure also depends on $n$. Analogous
toric compactifications of the form M$_{D-p}\times $T$^{p}$ were studied in 
\cite{R25}, and warped brane-like configurations were also discussed in both 
\cite{R24} and \cite{R25}.

It was shown in \cite{R24} that, in spite of the difference between Lovelock
theory and Einstein theory, the qualitative features of thermodynamic
stability of brane-like configurations in both theories are considerable
similar, although the higher order terms $\mathcal{R}^{n}$ can be seen to
contribute. For example, the thermodynamical analogue of Gregory-Laflamme
transition between black hole and black string configurations was discussed
in \cite{R24}. Extended string-like objects in Lovelock theory and their
thermodynamics were also discussed in \cite{H0412139,H0509102,H08011021}. We
discuss black hole thermodynamics in the next subsection.

\subsubsection*{Thermodynamics}

The purpose of this section is to describe the general aspects of black hole
thermodynamics in Lovelock theory. In fact, one of the most interesting
features of the Lovelock theory regards the thermodynamics of its black hole
solutions. This is because it is in the analysis of the black hole
thermodynamics where the substantial differences between Lovelock theory and
Einstein theory manifest themselves.

Pioneer works where the Lovelock black hole thermodynamics was discussed in
detail are references \cite{HMyers,HWhitt}; see also \cite{MyersOtros1,
MyersOtros2, MyersOtros3}. In \cite{R46}, Jacobson and Myers derived a close
expression for the entropy of these solutions in $D$ dimensions, and they
showed that the entropy of these black holes does not satisfy the area law,
but contains additional terms that are given by a sum of intrinsic curvature
invariants integrated over the horizon.

The thermodynamics of charged solutions was originally studied by Wiltshire
in Refs. \cite{Wiltshire, Wiltshire2}, while the thermodynamics of
topological black holes was studied more recently, in Refs. \cite%
{H9808067,T0011097,R41}. The study of charged topological black holes in
presence of cosmological constant was addressed in \cite{H0112045}, where
the most general solution of this type in EGB\ theory was obtained.
References \cite{H0202140,H0212092, H0302132} also analyze topological black
holes and their thermodynamics; see also \cite{R31,R51}.

The aim of this section is to discuss the more relevant thermodynamical
features of Lovelock solutions. To do this, we will consider again the
five-dimensional case (\ref{intervalo})-(\ref{sol}). Actually, besides it
represents a simple instructive example, the five-dimensional case is also
special in what concerns thermodynamical properties. It is the best example
to see that substantial differences between Lovelock gravity and Einstein
gravity exist.

It is easy to verify that the Hawking temperature associated to the solution
in $D=5$ with $\Lambda =0$ is given by

\begin{equation}
T=\frac{\hbar }{2\pi }\frac{r_{+}}{4\alpha +r_{+}^{2}}.  \label{T}
\end{equation}

Then, we see that, as expected, (\ref{T})\ behaves like the Hawking
temperature of a GR solution if the black hole is large enough, $%
r_{+}>>\alpha $, going like $T\simeq \hbar /8\pi r_{+}-\mathcal{O}(\alpha
/r_{+}^{3})$. On the other hand, temperature tends to zero for small values
of $r_{+}$, going like $T\simeq \hbar r_{+}/8\pi \alpha +\mathcal{O}%
(r_{+}^{3}/\alpha ^{2})$. This implies that the specific heat changes its
sign at length scales of order $r_{+}\sim \sqrt{\alpha }$, and a direct
consequence of this phenomenon is that five-dimensional Lovelock black holes
turn out to be thermodynamically stable, as they yield eternal remnants.
This can be easily verified by considering the rate of thermal radiation
which goes like $\partial _{t}M\sim -T^{5}r_{+}^{3}$, behaving like $dt\sim
-dr_{+}/r_{+}^{7}$ at short distances.


Nevertheless, it is worth pointing out that for dimension $D>5$ the
functional form of the temperature is substantially different from the case $%
D=5$, as it includes an additional term which is actually proportional to $%
(D-5)$. The general formula reads

\begin{equation}
T=\frac{\hbar }{4\pi }\frac{(D-3)r_{+}^{2}+2\alpha (D-5)}{4\alpha
r_{+}+r_{+}^{3}}.
\end{equation}%
which implies that, in $D>5$, the short distance limit is given by $T\simeq
(D-5)\hbar /8\pi r_{+}$, and the specific heat is then negative. This is the
reason why the thermodynamic behavior of higher dimensional
Einstein-Gauss-Bonnet black holes turns out to be more similar to that in
Einstein theory if $D\neq 5$. In general, eternal black holes arise in $%
D=2n+1$ dimensions if an $n^{\text{th}}$-order term $\mathcal{R}^{n}$ is
present in the action.

So, let us return to our instructive example of five dimensions. The entropy
associated to (\ref{T}) is given by

\begin{equation}
S=\frac{\mathcal{A}}{4G\hbar }+\mathcal{O}(\alpha r_{+})\sim
r_{+}^{3}+12\alpha r_{+},  \label{S}
\end{equation}%
from what we observe that black holes of Lovelock theory do not in general
obey the Bekenstein-Hawking area law. Actually, some particular solutions,
corresponding to topological black holes with flat horizon geometry $d\Sigma
_{3}^{2}=dx_{i}dx^{i}$, do obey the area law \cite{R29,R51}, but it is not
the case for spherically symmetric static solutions. A very interesting
discussion on the area law\footnote{%
In \cite{QWERT12} other corrections to area law were studied. The authors
thank S. Shankaranarayanan for pointing out this references to them.} is
that of Ref. \cite{R8}, where a version of the area law for symmetric
dynamical black holes defined by a future outer trapping horizon was
derived. There, the authors discussed the differences between the branches
of solutions with GR limit and those without it, and argue how for the
latter one still can define a concept of increasing dynamical entropy.

Notice that the second term in the right hand side of (\ref{S}) implies that
if $\alpha <0$ then the entropy turns out to be negative for sufficiently
small black holes\footnote{%
Refs. \cite{QWERT6,QWERT7} discuss related features. The authors thank S.
Odintsov for pointing out these references to them.}. This was discussed in 
\cite{R42}, where it was argued there that an additive ambiguity in the
definition of the entropy could be a solution for the negative entropy
contributions; see also the related discussion in \cite{T0011097}. In any
case, the theory for negative values of the coupling constant $\alpha $ is
somehow pathological in several respects. It not only gives negative
contributions to the entropy, but also ghost instabilities and strange
causal structure arise if $\alpha <0$. We will not consider the negative
values of $\alpha $ here.

Because of the current interest in black hole thermodynamics of higher order
theories, we consider convenient to mention that the entropy function
formalism, recently proposed by A. Sen \cite{Sen} within the context of the
attractor mechanism, works nicely for the case of Lovelock black holes. In
particular, this was recently studied in \cite{R47} for the case of EGB
black holes, and it was explicitly shown that (\ref{S}) is recovered by
analyzing the near horizon geometry. A rather general analysis was presented
in Ref. \cite{R48}. Very interesting discussions are those of Refs. \cite%
{QWERT9,QWERT10}.

The thermodynamic properties of topological black holes are also very
interesting; see for instance \cite{R37,R51}. As we already mentioned, it
can be shown that those black objects whose horizons are of zero curvature
do obey the area law for the entropy density. For instance, consider the
black brane geometry (\ref{Brane}), which is solution of the theory with
negative cosmological constant, $\Lambda <0$. It is straightforward to check
that the Hawking temperature of this solution is given by

\begin{equation}
T=\frac{\hbar }{6\pi }|\Lambda |r_{+},  \label{agree}
\end{equation}%
and that the area formula for the entropy density does hold in this special
case. Remarkably, identical expression for the temperature is obtained in
the particular case of the Chern-Simons theories of gravity, which we
discuss in the next subsection.

\subsubsection*{Chern-Simons black holes}

Now, let us move on, and analyze a very particular case of Lovelock theory
which exist in odd dimensions. This is the so-called Chern-Simons gravity
(CS), and can be thought of as a higher-dimensional generalization of the
Chern-Simons description of three-dimensional Einstein gravity \cite{Witten}%
. Basically, these theories are those particular cases of Lovelock
Lagrangian (\ref{L}) that admit a formulation in terms of a Chern-Simons
action. As we will discuss, these models are given by a very precise choice
of the set of coefficients $\alpha _{n}$.

To discuss CS gravity theories\footnote{%
It is worth pointing out that the CS theories we are referring to herein are
different to those discussed in Refs. \cite{QWERT3,QWERT4}.} it is
convenient\ to resort to the first-order formalism which, in spite of its
advantage, it is paradoxically avoided in physics discussions. So, let us
first review some basic notions: Consider the vielbein $e_{\mu }^{a}$, which
defines the metric as $g_{\mu \nu }=\eta _{ab}e_{\mu }^{a}e_{\nu }^{b}$,
where we are using the standard notation such that the greek indices $\mu
,\nu ,...$ correspond to the space-time while the latin indices $a,b,...$
are reserved for the tangent space. Now, consider the 1-form associated to
the vielbein, defined by $e^{a}=e_{\mu }^{a}dx^{\mu }$, and the
corresponding 1-form associated to the spin connection $\omega _{\mu }^{ab}$%
, defined by $\omega ^{ab}=\omega _{\mu }^{ab}dx^{\mu }$. These quantities
enable us to define the so-called curvature 2-form, which is given by 
\begin{equation*}
R^{ab}=d\omega ^{ab}+\omega _{\text{ }c}^{a}\wedge \omega ^{cb}=R_{~~\mu \nu
}^{ab}~dx^{\mu }\wedge dx^{\nu }\equiv \frac{1}{2}R_{~~\mu \nu
}^{ab}~(dx^{\mu }dx^{\nu }-dx^{\nu }dx^{\mu }),
\end{equation*}%
and is related to the Riemann tensor by $R_{\text{ }\beta \mu \nu }^{\alpha
}=\eta _{bc}e_{a}^{\alpha }e_{\beta }^{c}R_{\text{ \ }\mu \nu }^{ab}$. The
torsion-free condition is then given by 
\begin{equation*}
T^{a}=de^{a}+\omega _{b}^{a}\wedge e^{b}=0.
\end{equation*}

In this language, local Lorentz invariance of the theory is expressed in
terms of the covariant derivative 
\begin{equation}
\delta _{\lambda }\omega _{b}^{a}=d\lambda _{b}^{a}+\omega _{c}^{a}\wedge
\lambda _{b}^{c}-\omega _{b}^{c}\wedge \lambda _{c}^{a},\qquad \delta
_{\lambda }e^{a}=-\lambda _{b}^{a}e^{b},  \label{uno}
\end{equation}%
where $\lambda _{b}^{a}$ represent the parameters of the transformation.

The remarkable fact is that, for particular cases of the action (\ref{L}),
if the coupling constants are chosen appropriately, the theory exhibits an
additional local symmetry. For instance, if we consider the case $\Lambda =0$%
, such additional symmetry turns out to be given by the invariance of the
Lagrangian density under the gauge transformation%
\begin{equation}
\delta _{\lambda }e^{a}=d\lambda ^{a}+\omega _{b}^{a}\wedge \lambda
^{b},\qquad \delta _{\lambda }\omega _{b}^{a}=0.  \label{dos}
\end{equation}

That is, the CS theory possesses a local symmetry under gauge transformation 
$\delta _{\lambda }e_{\mu }^{a}=\partial _{\mu }^{a}\lambda +\omega _{b\mu
}^{a}\lambda ^{b}$, with $\lambda ^{a}$ being a parameter. This is actually
an off-shell local gauge symmetry of the theory (\ref{L}) that arises for
special choices of the coupling constants $\alpha _{n}$, as far as the
boundary conditions are also chosen in the appropriate way. Besides, it can
be easily verified that transformation (\ref{dos}), once considered together
with (\ref{uno}), satisfies the Poincar\'{e} algebra \thinspace $ISO(2,1)$,
and this is why these theories are usually referred as Poincar\'{e}%
-Chern-Simons gravitational theories \cite{T9601003}; see also \cite{Zanelli}
for an excellent introduction to Chern-Simons gravity.

So, let us specify which are the theories that possess the gauge symmetry
like\footnote{%
Notice that, as mentioned, (\ref{dos}) is the transofrmation that
corresponds to the case $\Lambda =0$. The analogous tranformation for the
case $l^{2}\neq 0$ takes a slightly different form, see \cite{Zanelli}.} (%
\ref{uno})-(\ref{dos}), namely the CS theories. To do this, first it is
convenient to rewrite the Lovelock Lagrangian. In the first-order formalism,
the Lovelock action corresponding to (\ref{L}) in $D=2t+1$ dimensions can be
written as

\begin{equation}
S=\int \varepsilon
_{a_{1}b_{1}a_{2}b_{2}...a_{t}b_{t}c}\bigwedge\nolimits_{n=1}^{t}\left(
R^{a_{n}b_{n}}+l_{n}^{-2}e^{a_{n}}\wedge e^{b_{n}}\right) \wedge e^{c}
\label{MN}
\end{equation}%
where $l_{n}^{-2}$ correspond to $t$ independent coefficients that are a
rearrangement of the coefficients $\alpha _{n}$. In (\ref{MN}), the
convention is such that the $t^{\text{th}}$ coupling $\alpha _{n=t}$ has
been set to $1$ (or, alternatively speaking, it has been absorbed in the
definition of the curvature $R^{ab}$), so that in this notation we have $%
|\Lambda |\sim \prod_{n=1}^{t}l_{n}^{-2}$, and $G^{-1}\sim
\sum\nolimits_{m=1}^{t}\prod_{n\neq m}l_{n}^{-2}$.

It is worth noticing that, in order to represent the most general form of (%
\ref{L}), the coefficients $l_{n}^{-2}$ in (\ref{MN})\ should be allowed to
take complex values. In fact, Lovelock action (\ref{L}) with real
coefficients $\alpha _{n}$ can correspond to (\ref{MN}) with imaginary $%
l_{n}^{-2}$. An example is given by the five-dimensional theory whose action
reads $S=\int \varepsilon _{abcdf}\left( R^{ab}+i\beta ^{2}\text{ }%
e^{a}\wedge e^{b}\right) \wedge \left( R^{cd}-i\beta ^{2}\text{ }e^{c}\wedge
e^{d}\right) \wedge e^{f}$, which leads to the particular form of (\ref{L})
where no Einstein-Hilbert contribution is present, but only the cosmological
constant and the Gauss-Bonnet term appear, with $\alpha /\Lambda \sim \beta
^{-4}$ for a real $\beta $.

The CS gravity theories, however, are given by real values of $l_{n}^{-2}$.
More precisely, CS\ theory correspond to the special case where the coupling 
$l_{n}^{2}$ in (\ref{MN}) combine to give only one value for the effective
cosmological constant $\Lambda _{\text{eff}}=\pm l^{-2}$. In terms of the
Lagrangian density (\ref{L}) this corresponds to taking the coupling
constants $\alpha _{n}$ to be $\alpha
_{n}=(-1)^{n+1}l^{2n-D}m!/((D-2n)(m-n)!n!)$ for$\ n>0$, while $\alpha _{0}$
is given by the cosmological constant $\Lambda =-\alpha _{0}/2\alpha _{1}$.
It is important to mention that (\ref{MN}) corresponds to the case of
negative cosmological constant, which yields the CS theory with the AdS$_{D}$
group (i.e. the group $SO(D-1,2)$) as the one that generates the gauge
symmetry. The case of positive $\Lambda $ is simply obtained by changing $%
l^{2}\rightarrow -l^{2}$, while the Poincar\'{e} invariant theory is
obtained through the Inonu-Winger contraction of (A)dS group; see \cite%
{Zanelli} for details. An example of Poincar\'{e} invariant CS is given by
the Lagrangian containing only the quadratic Gauss-Bonnet $\sqrt{-g}\mathcal{%
R}^{2}$ term in five dimensions, without the Einstein-Hilbert term and with $%
\Lambda =0$.

As it is well known, an example of the CS gravity theory is given by
three-dimensional Einstein theory, whose action\footnote{%
For simplicity here we have fixed the Newton constant according to $16\pi
G=1 $.},%
\begin{equation}
S=\int d^{3}x\ \mathcal{L}=\int d^{3}x\sqrt{-g}\left( R-2\Lambda \right) ,
\label{quince}
\end{equation}%
admits to be formulated as a CS theory. To see this, and then extend the
construction to higher dimensional cases, let us first point out that (\ref%
{quince}) can be written as follows,%
\begin{equation}
S=\int_{\text{M}_{3}}\varepsilon _{abc}(R^{ab}\wedge e^{c}-l^{-2}e^{a}\wedge
e^{b}\wedge e^{c}),  \label{Secasa}
\end{equation}%
with $\Lambda \sim l^{-2}$.

It turns out that (\ref{quince})-(\ref{Secasa}) admits to be formulated as a
CS theory \cite{Witten}\ for the groups $SO(2,2)$, $SO(3,1)$ and $ISO(2,1)$,
depending on whether the cosmological constant $\Lambda $ is negative,
positive or zero, respectively. To make contact with the usual form of the
CS action, let us introduce a ($D+1$)-dimensional 1-form $A^{ab}$ whose
indices run over $a,b=0,1,2,...,2t+1$ (recall $D=2t+1$), and its strength
field $F^{ab}=dA^{ab}+A_{\text{ }c}^{a}\wedge A^{cb}$, which are given by%
\begin{equation*}
A^{ab}=%
\begin{pmatrix}
\omega ^{ab} & e^{a}/l \\ 
-e^{b}/l & 0%
\end{pmatrix}%
,\ \ \ F^{ab}=%
\begin{pmatrix}
R^{ab}-l^{-2}e^{a}\wedge e^{b} & l^{-1}\left( de^{a}+\omega _{c}^{a}\wedge
e^{c}\right) \\ 
-l^{-1}\left( de^{b}+\omega _{c}^{b}\wedge e^{c}\right) & 0%
\end{pmatrix}%
.
\end{equation*}%
That is, $A^{ab}=\omega ^{ab}$ for $a,b=0,1,2,...,2t$, while $%
A^{aD}=-A^{Da}=e^{a}/l$ for $a=0,1,2...2t$. Analogously, $%
F^{ab}=R^{ab}-l^{-2}e^{a}\wedge e^{b}$ for $a,b=0,1,...2t$, while $%
F^{aD}=-F^{Da}=T^{a}/l$ for $a=0,1,2,...2t$.

Then, making use of these definitions, (\ref{quince})-(\ref{Secasa}) can be
alternatively expressed in its Chern-Simons form%
\begin{equation}
S=\int_{\text{M}_{3}}\text{Tr\ }(A\wedge dA+\frac{2}{3}A\wedge A\wedge A),
\label{SCS}
\end{equation}%
where the trace is over the indices $a,b$ that run from $0$ to $3$
(corresponding to $D=3$, i.e. $t=1$). Local symmetry under (\ref{uno}) and (%
\ref{dos}) is then gathered by gauge symmetry of (\ref{SCS}).

The next example we could consider is the five-dimensional one, which
corresponds to the Lovelock theory (\ref{L}) for the particular case $\alpha
_{0}\alpha _{2}=3/2$ (i.e. $\alpha \Lambda =-3/4$). Then, the action reads

\begin{equation}
S=\int d^{5}x\ \mathcal{L}=\int d^{5}x\sqrt{-g}(R+\frac{2}{l^{2}}-\frac{%
3l^{2}}{4}(R+R_{\mu \nu \alpha \beta }R^{\mu \nu \alpha \beta }-4R^{\mu \nu
}R_{\mu \nu }))
\end{equation}%
where $\Lambda =-l^{-2}$ and $\alpha _{2}=3/2\alpha _{0}=-3/4\Lambda
=3l^{2}/4$. This can be also written as%
\begin{equation}
S=\int_{\text{M}_{5}}\mathcal{\varepsilon }_{abcdf}\ (R^{ab}\wedge R^{cd}+%
\frac{2}{3l^{2}}R^{ab}\wedge e^{c}\wedge e^{d}+\frac{1}{5l^{4}}e^{a}\wedge
e^{b}\wedge e^{c}\wedge e^{d})\wedge e^{f}
\end{equation}%
and, again, it admits to be written in its Chern-Simons form 
\begin{equation}
S=\frac{1}{\kappa ^{2}}\int_{\text{M}_{5}}\text{Tr\ }(A\wedge \left(
dA\right) ^{\wedge 2}+\frac{3}{2}\left( A\right) ^{\wedge 3}\wedge dA+\frac{3%
}{5}\left( A\right) ^{\wedge 5})
\end{equation}

Actually, this structure goes on as $D$ increases, and it expands a whole
family of theories which, still being particular cases of Lovelock theory (%
\ref{L}), represent odd-dimensional field theories with local off-shell
symmetry under the (A)dS (or Poincar\'{e}) group.

Now, once we have introduced the theories, let us analyze their black hole
solutions. Going back to solution (\ref{intervalo}), and considering again
the five-dimensional case as an example, we observe that replacing the
Chern-Simons condition\footnote{%
It is helthy to consider the case $\alpha >0$ and $\Lambda <0$.} $\alpha
\Lambda =-3/4$ in the metric function (\ref{sol}) leads to a rather
different geometry, given by 
\begin{equation}
V^{2}(r)=\frac{r^{2}}{4\alpha }-\mathcal{M}\quad \quad \text{with}\quad
\quad \mathcal{M}+1=-\sigma \sqrt{M/\alpha }.  \label{sol2}
\end{equation}

This solution still may represent a black hole, provided $\mathcal{M}>0$,
with the horizon located at $r_{+}=2\sqrt{\mathcal{M}/\alpha }$. However,
this is a black hole of a different sort. In particular, it does not present
a limit where GR is recovered, and this can be understood in terms of the
condition $\alpha =-3/4\Lambda $ in the following way: While the
cosmological constant $\Lambda $ introduces an infrared cut-off (the length
scale $1/\sqrt{|\Lambda |}$) where the cosmological term dominates over the
Einstein-Hilbert term, the Gauss-Bonnet term introduces an ultraviolet
cut-off (the length scale $\sqrt{\alpha }$) where the quadratic terms
dominate. Therefore, the condition $\alpha =-3/4\Lambda $ basically states
that in Chern-Simons theory both length scales are of the same order, and
consequently there is no range where the Einstein-Hilbert term is the
leading one. This explains why there is no range where (\ref{sol2})
approaches Schwarzschild-Tangherlini solution. This asphyxia of the
Einstein-Hilbert term is a typical feature of Chern-Simons theories for $D>3$%
, where a unique free parameter $l^{2}$ appears in the action.

The Hawking temperature associated to black hole solution (\ref{sol2}) is
given by%
\begin{equation}
T=\frac{\hbar }{8\alpha \pi }r_{+}=\frac{\hbar }{6\pi }|\Lambda |r_{+},
\label{naitin}
\end{equation}%
which in turn agrees with (\ref{agree}), although now it corresponds to a
spherically symmetric solution. As it is well known \cite{BTZbis, BTZ} in $%
D=3$ formula (\ref{naitin}) agrees with the area law.

Certainly, solution (\ref{sol2}) is reminiscent of the Ba\~{n}%
ados-Teitelboim-Zanelli three-dimensional black hole (BTZ), which, after
all, also corresponds to a CS black hole. In fact, this is not a
coincidence, and regarding this, let us make a historical remark: It turns
out that, even though one could imagine that CS black holes (\ref{sol2})
were discovered as higher-dimensional extensions of the BTZ, the story was
precisely the opposite: In 1992, Ba\~{n}ados, Teitelboim and Zanelli
discovered the BTZ as a particular case of a family of Lovelock black holes
they were studying at that time \cite{BTZ4,BTZ2,BTZ3}.

The analogy between the BTZ\ black hole and those solutions for
higher-dimensional CS theories was discussed in detail in \cite{AFG}. In
particular, it was emphasized there that five-dimensional solution (\ref%
{sol2}) shares several properties with its three-dimensional analogue. For
instance, it is the case of their thermodynamics properties, which, after
all, are actually encoded in the function $V^{2}(r)$. This is also why all
CS black holes have infinite lifetime.

Notice that the parameter $\mathcal{M}$ in Eq. (\ref{sol2}) plays the role
that the mass $M$ plays in the BTZ solution. Also, as in the
three-dimensional case, the Anti-de Sitter space is obtained for a
particular value of this parameter, namely $\mathcal{M}=-1$, and a naked
singularity is developed for the range $-1<\mathcal{M}<0$.

In \cite{CTZ} the CS black holes and their dimensional extensions were
exhaustively studied, together with their topological and charged
extensions. There, a very interesting class of black holes was found by
considering the particular choice of coefficients that leads to the $(2t+1)$%
-dimensional CS theory, but dimensionally extending the action from $D=2t+1\ 
$to $D\geq 2t+1$. The metrics of such solutions are given by replacing the
constant $\mathcal{M}$ in (\ref{sol2}) by the quantity $1-\mathcal{M}%
r^{(2t+1-D)/t}$. A further generalization of the solutions of \cite{CTZ}
would be given by adding a volume term to the gravitational action, which in
turn corresponds to shifting the coupling $\alpha _{0}\rightarrow \alpha
_{0}+\delta \Lambda $ but keeping the rest of $\alpha _{n>0}$ tuned as they
are in the ($2t+1$)-dimensional CS theory, given in terms of the length
scale $l^{2}$. The solution for this case is given by replacing the constant 
$\mathcal{M}$ in (\ref{sol2}) by a term $1-\left( r^{2t}+\lambda r^{2t}+%
\mathcal{M}r^{2t+1-D}\right) ^{1/t}/l^{2}$, where $\lambda +1\sim \delta
\Lambda /\alpha _{0}$. These black holes do have a GR limit since now the
cosmological length scale can be pushed away by choosing $\delta \Lambda $
appropriately.

It is also important to mention that black hole solution (\ref{sol2}) is
also a solution of the CS theory with torsion \cite%
{Giacomini,Giacomini2,QWERT11}.

The solutions of Chern-Simons theory are very special ones, and this is due
to the fact that for that specific choice of the coupling constants $\alpha
_{n}$ the equations of motion of Lovelock theory somehow degenerate. In
particular, it is remarkable that the obstruction imposed by Birkhoff-like
theorems does not hold for CS theories.

\subsubsection*{A word on spinning black holes}

Before concluding this section, a word on the spinning black hole case:\ The
problem of finding a rotating solution in Einstein-Gauss-Bonnet gravity,
which would generalize the Kerr's spinning black hole of GR, is a hard and
still unsolved problem. Recently, it was proven in \cite{ROT2} that the
Kerr-Schild ansatz does not work in Lovelock theory (except for very special
cases as Einstein theory and Chern-Simons theory), and this manifestly shows
how difficult this classical problem is.

Nevertheless, despite the difficulty, some advances in this area were recently achieved: In \cite{ROT2} an exact analytic rotating solution was found 
for Chern-Simons gravity in five dimensions. This Einstein-Gauss-Bonnet solution, however, does not present a horizon, and thus it does not represent a 
black hole. Nevertheless, the numerical analysis of \cite{H08011021} supports the idea that rotating solutions actually exist. Besides, approximated 
analytic solutions at first order in the angular momentum parameter were found in \cite{H08011021ZZZ}. Other solutions are known which represent 
rotating flat branes; these are a simple extension of topological black holes with $k=0$.

Despite these recent advances, the problem of finding an exact analytic rotating black hole solution in Lovelock theory still remains an open 
problem.

\section{Including boundary terms}

In this section, we will discuss other constructions which, locally,
coincide with the Deser-Boulware spherically symmetry metric.

\subsubsection*{Wormholes}

The next class of solutions we would like to discuss is a class of vacuum
solutions of Lovelock theory which represents wormhole geometries that
connect two disconnected asymptotic regions of the space-time. Recently,
several examples of such solutions were found \cite{HBhawal,GGGW,
GravanisWillison3, R5,R7,H07102041,COT}, describing vacuum wormholes with
different asymptotic behaviors, and in different number of dimensions. So,
the first question we might ask is: why do wormholes exist in Lovelock
theory?

The main reason why vacuum wormholes exist in a theory like (\ref{L}) is
actually simple, and it can be heuristically explained as follows: Consider
the equations of motion corresponding to Lagrangian (\ref{L}), which can be
always written as%
\begin{equation}
R_{\mu \nu }-\frac{1}{2}Rg_{\mu \nu }+\Lambda g_{\mu \nu }-T_{\mu \nu }=0
\label{eco}
\end{equation}%
where the higher order terms act as an effective stress tensor that here we
denoted $T_{\mu \nu }$. In the case of EGB theory it reads%
\begin{equation*}
\frac{1}{\alpha }T_{\mu \nu }=\frac{1}{2}g_{\mu \nu }\left( R_{\rho \sigma
\alpha \beta }R^{\rho \sigma \alpha \beta }-4R_{\alpha \beta }R^{\alpha
\beta }+R^{2}\right) -2RR_{\mu \nu }+4R_{\mu \rho }R_{\nu }^{\rho
}+R_{\alpha \beta }R_{\ \ \mu \nu }^{\alpha \beta }-2R_{\mu \alpha \beta
\rho }R_{\nu }^{\ \alpha \beta \rho },
\end{equation*}%
where, as usual, $\alpha =\alpha _{2}/\alpha _{1}$, $2\Lambda =-\alpha
_{0}/\alpha _{1}$. The key point is that this effective stress tensor $%
T_{\mu \nu }$, thought of as a kind of matter contribution, can be shown to
violate the energy conditions for $\alpha $ large enough. Actually, this
does not represent an actual problem from the conceptual point of view since
this "matter" is actually made of pure gravity. However, a consequence of
this violation of the energy conditions is that Eqs. (\ref{eco})\ allow the
existence of vacuum wormhole solutions at scales of order $\sqrt{\alpha }$,
unlike the case of GR, where solutions of this sort require the
consideration of \textit{exotic matter}.

Furthermore, there is a second reason for such curious solutions to exist in
Lovelock theory. As mentioned above, when the coefficients $\alpha _{n}$ in (%
\ref{L}) correspond to the CS theory, the space of solutions experiments an
unusual enhancement, which translates into a large degeneracy of the metric
of spaces with enough symmetry. Roughly speaking, for such particular cases,
Lovelock theory is somehow degenerated enough to admit metric with very
special properties, and wormholes are some of them.

Nevertheless, here we will focus our attention on wormhole solutions that
exist in five-dimensional EGB theory without requiring the coefficients $%
\Lambda $ and $\alpha $ to be those that correspond to CS theory. Therefore,
the existence of such solutions, regarded as an anomaly, is ultimately
attributed to the issue of the energy conditions mentioned above.

\subsubsection*{Junction conditions}

The particular configurations we will consider are the so-called thin-shell
wormholes, which correspond to connecting two regions of the space through a
codimension-one hypersurface that plays the role of the wormhole throat. For
such a geometry to be constructed, we have to make use of the junction
conditions of the EGB theory \cite{MyersBoundary,GGGW}. In particular, we
will consider the configuration of two Boulware-Deser spaces connected
through a hypersurface on which the induced stress-tensor vanishes. Such
geometries are not possible in GR, where wormholes require the energy
conditions to be violated on the thin-shell. However, in Lovelock theory,
and because of the higher order terms, spherically symmetric vacuum
wormholes with positive mass can be constructed, as shown by Gravanis and
Willison in \cite{GravanisWillison3}. Let us review the procedure here.

Let $\Sigma $ be a four-dimensional timelike orientable hypersurface of
codimension one, whose normal vector is denoted by $n^{\mu }$. Suppose $%
\Sigma $ separates two regions of the space, which we call $\mathcal{M}_{I}$
and $\mathcal{M}_{II}$. Then, junction conditions read%
\begin{equation}
\left\langle K_{ij}-Kh_{ij}\right\rangle _{\Sigma }+2\alpha \left\langle
3J_{ij}-Jh_{ij}+2P_{iklj}K^{kl}\right\rangle _{\Sigma }=8\pi S_{ij}
\label{JC}
\end{equation}%
where $\left\langle X\right\rangle _{\Sigma }$ denotes the jump of the
quantity $X$ across the hypersurface $\Sigma $, which means $\left\langle
X\right\rangle _{\Sigma }=X_{|_{II}}\pm X_{|_{I}}$, where the sign $\pm $
depends on the relative orientation of the regions. Above, tensor $S_{ij}$
represents the induced stress-tensor on the hypersurface $\Sigma $, in
complete analogy with the Israel junction conditions in Einstein theory. In
fact, we see that the first two terms in (\ref{JC}) actually correspond to
the Israel junction conditions constructed with the extrinsic curvature $%
K_{i}^{j}$ and its trace $K$. In addition, the junction conditions
corresponding to the EGB theory contains contributions cubic in the
extrinsic curvature\footnote{%
See \cite{Charmonius}\ for a recent review. See also \cite{QWERT13} where
boundary terms in odd-dimensions are discussed.},%
\begin{equation}
J_{ij}=\frac{1}{3}%
(K_{kl}K^{kl}K_{ij}+2KK_{ik}K_{j}^{k}-K^{2}K_{ij}-2K_{ik}K^{kl}K_{lj}),
\end{equation}%
and also contributions that involve the Riemann curvature tensor of the
hypersurface 
\begin{equation}
P_{ijkl}=R_{ijkl}+R_{jk}h_{il}-R_{jl}h_{ik}+R_{il}h_{jk}-R_{ik}h_{jl}+\frac{1%
}{2}Rh_{ik}h_{jl}-\frac{1}{2}Rh_{il}h_{jk}.
\end{equation}

The notation used here is such that latin indices \thinspace $i,j,k,l$ refer
to coordinates on the four-dimensional hypersurface that separate the two
five-dimensional regions of the space. The induced metric is denoted by $%
h_{ij}$. It is worth mentioning that in Ref. \cite{GGGW} the junction
conditions were studied in the most general case, including the case of
space-like junctures, which corresponds to a cosmological-type geometries
that experiment a change of behavior at a given time characterized by the
hypersurface $\Sigma $. It was pointed out by H. Maeda that this kind of
space-like junction conditions could be used to construct regular black hole
solutions by means of geometric surgery procedure inside the black hole
horizon.

Here we will be mainly concerned with static spherically symmetric
geometries, and, besides, with spherically symmetric boundary conditions.
That is, we will consider the time-like hypersurface $\Sigma $ that
separates the two regions of the space to be located at fixed radial
coordinate $r=a(\tau )$, and the system of coordinates we will parameterize
the three angular directions $\phi _{1},\phi _{2},\phi _{3}$ of the junction
hypersurface, and the proper time $\tau $ of an observer on $\Sigma $.

Then, we introduce the metrics 
\begin{equation}
ds_{I,II}^{2}=-K_{I,II}^{2}(r)\,dt^{2}+K_{I,II}^{-2}(r)dr^{2}+r^{2}d\Sigma
_{3}^{2}\,,  \label{bulk L}
\end{equation}%
on each region $\mathcal{M}_{I}$ and $\mathcal{M}_{II}$, and the two regions
join at $r=a(\tau )$. Since here we consider vacuum solutions, $K_{I}^{2}(r)$
and $K_{I}^{2}(r)$ are given by (\ref{Siete}) (or by (\ref{sol}) in the case 
$k=1$). In general, there is no reason for the mass parameters $M_{I,II}$ of
the two regions to be equal, and the same happens with the choice of the
branches $\sigma _{I,II}=\pm 1$. Moreover, the orientation of $\mathcal{M}%
_{I}$ and that of $\mathcal{M}_{II}$ with respect to the normal vector $%
n^{\mu }$ are also independent one on each other, and we will take this
degree of freedom into account by introducing the variables $\eta _{I}$ and $%
\eta _{II}$ which indicate whether in each region the radial coordinate $%
r_{I,II}$ is parallel ($\eta _{I,II}=1$) or anti-parallel ($\eta _{I,II}=-1$%
) to $n^{\mu }$. Therefore, wormhole-like geometry corresponds to the
orientation $\eta _{I}\eta _{II}=-1$, while the standard shell-like\
geometry corresponds to the case $\eta _{I}\eta _{II}=+1$. The freedom in
choosing the parameters $M$, $\sigma $, $\eta $ independently in each region
allows for a wide class of solutions. The whole catalog was recently studied
in \cite{GGGW}.

The metric on $\Sigma $ induced from region $\mathcal{M}_{I}$ is the same as
the one induced from region $\mathcal{M}_{II}$, and is given by%
\begin{equation}
d\hat{s}^{2}=-d\tau ^{2}+a_{(\tau )}^{2}d\Sigma _{3}^{2}\,,
\label{4-geometry}
\end{equation}%
where, according to (\ref{Siete}), $d\Sigma _{3}^{2}$ will be chosen to be
the line element of a 3-manifold with (intrinsic) curvature $k=+1,-1,0$,
i.e. it is a unit sphere, a hyperboloid or flat space respectively. The
hypersurface $\Sigma $ is the world-volume of the juncture where regions $\mathcal{M}_{I}$ and $\mathcal{M}_{II}$ join. 

\begin{figure}[tbp]
\begin{center}
\includegraphics[height=0.50\textwidth,  angle=270]{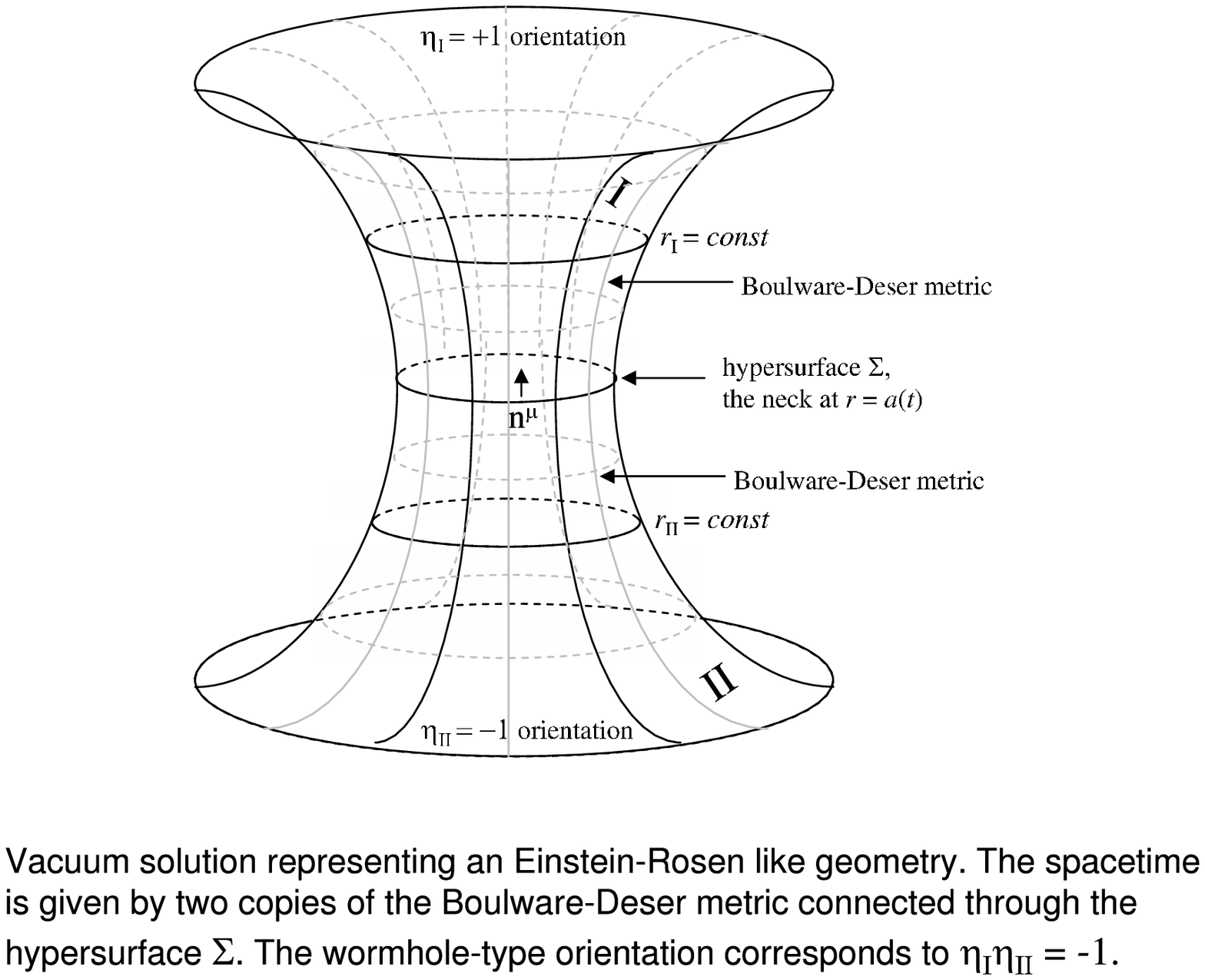}  
\\[0pt]
\end{center}
\caption{Einstein-Rosen bridge geometry as a vacuum solution.}
\end{figure}

To see whether such a wormhole-like (or shell-like) configuration is
possible in vacuum, we have to solve junction conditions (\ref{JC}) with $%
S_{i}^{j}=0$. To do this we first need to compute the components of the
intrinsic curvature. These are given by

\begin{equation*}
K_{\phi _{i}}^{\phi _{i}}=\frac{1}{a}(V^{2}(a)+\left( \partial _{\tau
}a\right) ^{2})^{1/2},\qquad K_{\tau }^{\tau }=(\partial _{\tau }^{2}a+\frac{%
1}{2}\partial _{r}V^{2}(a))(V^{2}(a)+\left( \partial _{\tau }a\right)
^{2})^{-1/2}
\end{equation*}%
with $i=1,2,3$. This also yields%
\begin{equation*}
3J_{\tau }^{\tau }-J=\frac{2}{a^{3}}(V^{2}(a)+\left( \partial _{\tau
}a\right) ^{2})^{3/2},\quad 3J_{\phi }^{\phi }-J=\frac{2}{a^{2}}%
(V^{2}(a)+\left( \partial _{\tau }a\right) ^{2})^{1/2}(\partial _{\tau
}^{2}a+\frac{1}{2}\partial _{r}V^{2}(a)).
\end{equation*}%
On the other hand, the components of Riemann tensor $R_{\ kl}^{ij}$ and
those of $P_{\ kl}^{ij}$ are%
\begin{equation*}
R_{\ \tau \phi _{i}}^{\tau \phi _{i}}=P_{\ \phi _{i}\phi _{j}}^{\phi
_{i}\phi _{j}}=\frac{1}{a}\partial _{\tau }^{2}a,\quad \quad R_{\ \phi
_{i}\phi _{j}}^{\phi _{i}\phi _{j}}=P_{\ \phi _{i}\tau }^{\phi _{i}\tau }=%
\frac{1}{a^{2}}(1+\left( \partial _{\tau }a\right) ^{2}).
\end{equation*}

Putting all this together, we can evaluate the junction conditions (\ref{JC}%
) in vacuum. The two independent equations read%
\begin{equation*}
(\eta _{I}V_{I}(a_{0})-\eta _{II}V_{II}(a_{0}))(a_{0}^{2}+\frac{4\alpha }{3}%
(3k-V_{I}^{2}(a_{0})-V_{II}^{2}(a_{0})-\eta _{I}\eta
_{II}V_{I}(a_{0})V_{II}(a_{0})))=0,
\end{equation*}%
\begin{equation*}
(\eta _{I}V_{I}^{-1}(a_{0})-\eta _{II}V_{II}^{-1}(a_{0}))(k-\frac{\Lambda
a_{0}^{2}}{3}-\eta _{I}\eta _{II}V_{I}(a_{0})V_{II}(a_{0}))=0.
\end{equation*}

For the wormhole orientation, $\eta _{I}\eta _{II}=-1$, and for the
symmetric case $V_{I}^{2}(a)=V_{II}^{2}(a)$, these equations take the simple
form%
\begin{equation}
V^{2}(a_{0})=\frac{3}{4\alpha }a_{0}^{2}+3k,\qquad V^{2}(a_{0})=\frac{%
\Lambda a_{0}^{2}}{3}-k.  \label{RR}
\end{equation}

From these equations we see that the radius of the throat of the wormhole is
given by%
\begin{equation}
a_{0}^{2}=\frac{12\alpha k}{\alpha \Lambda -9/4},  \label{mmm}
\end{equation}%
and from this we can also calculate the mass of the wormhole easily. Eq. (%
\ref{mmm}) implies that, in the case of the spherically symmetric wormhole ($%
k=1$), we need $\Lambda \alpha >9/4$ for the wormhole to exist. Then,
provided $\Lambda \alpha $ is of order one, the radius of the wormhole
throat is or order $a_{0}\sim \sqrt{\alpha }$. Besides, we should ask the
throat to be located outside the horizon, namely $a_{0}>r_{+}$. It is
remarkable that all these conditions can be satisfied \cite{GGGW} for
positive values of $\alpha $, $k$, $M$ and $\Lambda $. However, it is worth
mentioning that spherically symmetric wormhole solutions only exist if at
least one of the two regions $\mathcal{M}_{I,II}$ corresponds to the branch $%
\sigma =+1$ in (\ref{sol}).

More remarkable is the fact that the analysis of the dynamic case $a=a(\tau
) $ follows straightforward. When $\partial _{\tau }a\neq 0$, equations $%
S_{\tau }^{\tau }=0$ and $S_{\phi _{i}}^{\phi _{i}}=0$ are not linearly
independent, and it is sufficient to solve the first of them. Considering $%
k=1$, we get%
\begin{equation}
\left( \partial _{\tau }a\right) ^{2}+W(a)=0\quad \text{with}\quad W(a)=%
\frac{1}{4\alpha }a^{2}-\frac{\sigma }{2}\sqrt{\frac{a^{4}}{16\alpha ^{2}}%
(1+4\Lambda \alpha /3)+\frac{MG}{\alpha }}+1,
\end{equation}%
which has the form of a one-dimensional dynamic equation of motion
constrained by the vanishing energy condition. Notice that (\ref{RR}) is
recovered by demanding $W(a)=0$. Notice also that the effective potential $%
W(a)$ has negative derivative, and for large values of $a$ it goes like $%
W(a)\simeq a^{2}(2-\sqrt{(1+4\Lambda \alpha /3)})/8\alpha <0$. The effective
potential $W(a)$ can be positive and of positive derivative for
non-symmetric wormhole configurations.

Summarizing, we have just seen that spherically symmetric (microscopic)
thin-shell wormholes in vacuum are admitted as solutions of the
five-dimensional Lovelock theory. These solutions are allowed by additional
terms arising in the junction conditions of the EGB theory.

It is worth mentioning that the one we discussed here is not the only class
of wormhole-like solutions that exists in Lovelock theory. For instance, in 
\cite{R5,COT} a static wormhole solution for gravity in vacuum was found for
CS gravity in arbitrary (odd) dimensions $D=2t+1\geq 5$. This wormhole
connects two asymptotic regions whose respective boundaries are locally
given by R $\times $ $S^{1}\times $ H$_{d-3}$.

Besides, $D$-dimensional static wormhole solutions of the EGB theory were
also studied in \cite{R7}, and explicit wormhole solutions respecting the
energy conditions in the whole spacetime were found for the case $\alpha >0$%
. The asymptotic behavior of these solutions is given \ by R $\times $ H$%
_{d-2}$.

\subsubsection*{Naked singularities}

As we have seen in the previous sections, there are many features of
Lovelock solutions that are not present in GR. Eternal black holes and
wormholes are remarkable examples. Another example is the existence of
positive mass solutions with naked singularities\footnote{%
For a discussion on the formation of naked singularities, see\cite{R14,R11}.}%
. In fact, naked singularities appear in all the catalog of solutions, for
both spherically symmetric and extended objects, for both solutions with a
suitable GR limit and solutions without it. But, what kind of naked
singularities are these? For instance, we could ask whether these are stable
under gravitational perturbations \cite{Gibbons, DottiNegative}; or whether
these turn out to be "bad" singularities when probed with wave functions 
\cite{HorowitzMarolf}.

Regarding the question about the stability, this issue was studied recently
within the framework of the Kodama-Ishibashi formalism, and some evidence of
instabilities was found \cite{DottiBrasil}. On the other hand, here we will
address the second question, the one about how these naked singularities
look like when analyzed with quantum probes. To do this we will employ the
method developed by Horowitz and Marolf in Ref. \cite{HorowitzMarolf}, based
on the pioneer work of Wald \cite{Wald2}. The basic idea es the following:
Unlike what happens in the classical regime, where a singular space is
defined by the concept of geodesic incompleteness, in the quantum mechanical
regime the singular character of the space-time is defined in terms of the
ambiguity in the definition of the Hamiltonian evolution of wave functions
on it \cite{HorowitzMarolf}. More specifically, the singular nature of a
given space is determined in terms of the ambiguity when trying to find a
self-adjoint extension of the Hamiltonian operator to the whole space. When
such self-adjoint extension exists and is unique, then it is said that the
space is quantum mechanically regular, in spite of the singularities it
might present at classical level. Notice that this is not matter of
deforming the space or somehow resolving it, but it is rather a
reconsideration of what is the relevant physical dynamics on it. In fact, a
space can be classically singular but still regular when it is analyzed with
quantum probes.

Here, we will apply the concept of quantum probes to the singular solutions
of Lovelock theory discussed above. But, first, let us review the method
developed in \cite{HorowitzMarolf, IshibashiWald}. Consider the quantum
dynamics of a scalar field $\varphi $\ on the spherically symmetric space (%
\ref{intervalo}), which is governed by the Klein-Gordon equation%
\begin{equation}
\left( \nabla _{\mu }\nabla ^{\mu }-m^{2}-2\xi R\right) \varphi =0.
\label{KG}
\end{equation}%
This equation can be written as follows%
\begin{equation}
\partial _{t}^{2}\varphi +\mathcal{H}^{2}\varphi =0,\quad \text{with}\quad 
\mathcal{H}^{2}=-V_{(r)}\nabla ^{i}\left( V_{(r)}\nabla _{i}\varphi \right)
+V_{(r)}^{2}m^{2}\varphi +2V_{(r)}^{2}\xi R\varphi  \label{adove}
\end{equation}%
where $\nabla ^{i}$ is the covariant derivative on the spacelike
hypersurfaces defined by constant $t$ foliations, and where the metric
function $V^{2}(r)$ is given by (\ref{sol}). The piece $V_{(r)}\nabla
^{i}\left( V_{(r)}\nabla _{i}\varphi \right) $ in (\ref{adove}) involves the
Laplacian operator on the unitary 3-sphere, whose eigenvalues are known to
be given by $-l(l+2)$ with positive integers $l=0,1,2,3,...$

Now, equation (\ref{adove}) can be written in its Schr\"{o}dinger-like form,
schematically, 
\begin{equation*}
i\partial _{t}\varphi =\mathcal{H}\varphi ,
\end{equation*}%
and then the problem to deal with is to decide whether the Hamiltonian
operator $\mathcal{H}$ admits a unique self-adjoint extension in spite of
the fact the space is singular at the origin $r=0$. As mentioned, in the
quantum mechanical context the existence of singularity is associated to the
non-existence of a unique self-adjoint extension of the Hamiltonian operator
rather than to a geodesical completeness. Then, the problem of determining
whether the space is regular is translated into the problem of verifying
whether $\mathcal{H}^{2}$ admits a unique self-adjoint extension $\mathcal{H}%
_{E}^{2}$. If such extended operator exists, then the Hamiltonian evolution
of the wave function in this space would be given by%
\begin{equation*}
\varphi (t)=\exp (-it\text{ }\mathcal{H}_{E})\text{ }\varphi (0),
\end{equation*}%
and it would be well-defined.

It turns out that a sufficient condition for $\mathcal{H}_{E}^{2}$ to exist
and be unique is that at least one of the solutions of the differential
equation

\begin{equation}
\partial _{r}^{2}\phi _{(r)}+\partial _{r}\log \left(
r^{3}V_{(r)}^{2}\right) \partial _{r}\phi _{(r)}-V_{(r)}^{-2}\left(
r^{-2}l(l+2)-m^{2}+\xi R\pm iV_{(r)}^{-2}\right) \phi _{(r)}=0
\label{differential}
\end{equation}%
fails to be of finite norm near the origin for any value of $l$ and for any
of the two possible signs $\pm $ in (\ref{differential}); see \cite%
{HorowitzMarolf} for details. In other words, for the space to be considered
regular quantum mechanically it is necessary to see that at least one
solution $\phi $ to (\ref{differential}) is non-normalizable around the
origin. This criterion strongly depends on which norm $||\phi ||$ is
considered.

The well-posedness of an initial value problem requires not only the
existence and uniticity of conditions, but also continuous dependence of
solutions on initial data. Then, the norm $||\phi ||$ to be considered
should select a the function space that fulfills these requirements. A
sensitive norm in this sense is the Sobolev norm \cite{HosoyaIshibashi}.

To see how the method works in the case we are interested in, let us
consider again the five-dimensional Boulware-Deser space (\ref{intervalo})-(%
\ref{sol}). The branch $\sigma =+1$ of this space presents a naked
singularity for all positive values of $M$, while the branch $\sigma =-1$
only presents naked singularities within the range $0<M<\alpha $. Then, let
us solve the wave equation for these spaces. To analyze the solutions of (%
\ref{differential}) near the singular point $r=0$ it is convenient to write
this equation as $\partial _{r}^{2}\phi +r^{-1}p_{(r)}\partial _{r}\phi
+r^{-2}q_{(r)}\phi =0$, with $p(r)$ and $q(r)$ being two functions analytic
at the origin. \noindent This is a Fuchsian equation and so it admits
solutions with the form $\phi (r)=r^{\eta }f(r)$ for certain analytic
function $f(r)$ and a complex number $\eta $ that is known to solve the
indicial equation $\eta ^{2}+(p_{(r=0)}-1)\eta +q_{(r=0)}=0$. Then,
replacing (\ref{sol}) in (\ref{differential}) we find $p_{(r=0)}=3$, $%
q_{(r=0)}=-l(l+2)/(1+\sigma \sqrt{M/\alpha })$, and two independent
solutions to (\ref{differential}) are then given by the two values of $\eta $
that solve $\left( \eta +1\right) ^{2}=1+l(l+2)/(1+\sigma \sqrt{M/\alpha })$%
. Therefore, we find that one of the solutions to (\ref{differential})
always diverges at least as rapidly as $|\phi |^{2}\simeq r^{-2}$, and so it
fails to be integrable with respect to the Sobolev norm.

Summarizing, there exists a unique self-adjoint extension $\mathcal{H}%
_{E}^{2}$, from what we conclude that five-dimensional Boulware-Deser metric
turns out to be regular when tested by quantum probes. It is remarkable that
the positive (but small) mass solutions of five-dimensional black holes are
in a sense regular quantum mechanically, despite the naked curvature
singularity they exhibit at the origin.

Before concluding, we wish to make a remark about the consistency of
studying naked singularities in this way. Actually, one could wonder whether
probing naked singularities in a theory with a finite higher curvature
expansion makes sense or not. For instance, in string inspired models, as
soon as one approaches the singularity, neglecting higher order corrections
seems to be impossible since higher and higher order terms start to dominate
as we go close enough to the singularity. However, let us argue here that,
even though this is true, this is not necessarily an obstruction for testing
singularities with quantum probes up to certain order in the higher
curvature expansion. Let us be reminded of
what we do when we solve the Schr\"{o}dinger equation for the Coulombian
potential (e.g. In fact, the
analogy with the hydrogen atom in quantum mechanics is quite good since 
such problem also corresponds to solving a
wave function equation in presence of a central potential whose classical
counterpart breaks down at the origin). In quantum mechanics, even 
though
the Coulombian potential diverges at the origin, we know that the quantum
problem still makes sense, and we do solve the wave equation without
complaining about the fact that other corrections to the potential (e.g.
effective screening due to quantum effects, or short distance corrections to
the Coulombian potential) could in principle appear at very short distances.
Heuristically speaking, what one really has to do to make sure the whole
procedure makes sense is comparing the typical size of the wave packet with
the length scale where the terms that were neglected would dominate. For
example, above we were dealing with the EGB action, and the terms $\mathcal{R%
}^{3}$ were certainly neglected, and so the analysis carried out could still
make sense as long as the Compton length of the wave packet is small enough
in comparison with the length scale imposed by the coupling constant $\alpha
_{n}$ with $n>2$, and provided the fact higher curvature terms act as a
perturbation.

For some particular models where the couplings $\alpha _{n}$ are given in
terms of the same fundamental scale (like the models inspired in string
theory where the scale is given by $l_{s}^{2}\sim \alpha ^{\prime }$) the
story could be a little more subtle, and so the argument above would not be
valid in general. Nevertheless, it is likely the case that higher
order terms would contribute by smoothing out the singularity even more,
although not necessarily resolving it in a classical sense.

\subsection*{Acknowledgement}

This work was supported by UBA, CONICET, and ANPCyT, through grants UBACyT
X861, PIP6160, PICT34557. Conversations with M. Aiello, A. Anabal\'{o}n, E.
Ay\'{o}n-Beato, M. Ba\~{n}ados, C. Bunster, F. Canfora, G. Dotti, R. Ferraro, A. Garbarz, M. Hassa\"{\i}ne, D. 
Hofman, A. Giacomini, J.
Giribet, M. Kleban, D. Mazzitelli, R. Olea, J. Oliva, J. Saavedra, and D.
Tempo are acknowledged. The authors specially thank E. Gravanis and S.
Willison for collaborations in the subject, and they are grateful to J. Edelstein, M.
Leston, H. Maeda, R. Troncoso, and J. Zanelli for reading the manuscript and
making very important remarks. The authors also thank the members of the
CCPP at New York University for the hospitality. C.G. thanks the people of
the Brandeis Theory Group of Brandeis University.


\begin{thebibliography}{999}
\bibitem{Lovelock} D. Lovelock, \textit{The Einstein tensor and its
generalizations,} J. Math. Phys. \textbf{12} (1971) 498.

\bibitem{Lovelock2} D. Lovelock, \textit{The four-dimensionality of space
and the einstein tensor}, J. Math. Phys. \textbf{13} (1972) 874.

\bibitem{GrossWitten} D. Gross and E. Witten, {\it Superstrings 
modifications of Einstein's equations}, Nucl. Phys. {\bf B277} (1986) 1.

\bibitem{fR} H. Schmidt, \textit{Fourth order gravity: Equations, history,
and applications to cosmology}, Int. J. Geom. Meth. Mod. Phys. \textbf{4}
(2007) 209, [arXiv:gr-qc/0602017].

\bibitem{DeserTekin} S. Deser and B. Tekin, \textit{Shortcuts to high
symmetry solutions in gravitational theories}, Class. Quant. Grav. 
\textbf{20} (2003) 4877, [arXiv:gr-qc/0306114].

\bibitem{JP} R. Jackiw and S.-Y. Pi, \textit{Chern-Simons modification of
General Relativity}, Phys. Rev. D68 (2003) 104012, [arXiv:gr-qc/0308071].

\bibitem{YunesReview} S. Alexander and N. Yunes, {\it Chern-Simons Modified General Relativity}, to appear in 
Phys. Repts., [arXiv:0907.2562].

\bibitem{MTh} E. Witten, \textit{String theory dynamics in various dimensions%
}, Nucl. Phys. B443 (1995) 85, [arXiv:hep-th/9503124].

\bibitem{TSE} A. Tseytlin, \textit{$R^4$ terms in 11 dimensions and
conformal anomaly of (2,0) theory}, Nucl. Phys. \textbf{B584} (2000) 233,
[arXiv:hep-th/0005072].

\bibitem{GruzinovKleban} A. Gruzinov and M. Kleban, \textit{Causality
Constrains Higher Curvature Corrections to Gravity}, [arXiv:hep-th/0612015].

\bibitem{Minasian} I.Antoniadis, S. Ferrara, R. Minasian and K. Narain, {\it $R^4$ Couplings in M and Type II 
Theories}, Nucl. Phys. {\bf B507} (1997) 571, [arXiv:hep-th/9707013v2].

\bibitem{Minasian2} S. Ferrara, R. Khuri and R. Minasian, {\it M-Theory on 
a Calabi-Yau Manifold}, Phys. Lett. 
{\bf B375} (1996) 81, [arXiv:hep-th/9602102].

\bibitem{Strominger} M. Guica, L. Huang, W. Li and A. Strominger, \textit{R%
\symbol{94}2 Corrections for 5D Black Holes and Rings}, JHEP \textbf{0610}
(2006) 036, [arXiv:hep-th/0505188].

\bibitem{DEGB1} P. Kanti, N. Mavromatos, J. Rizos, K. Tamvakis and E. Winstanley, {\it Dilatonic Black Holes in Higher Curvature String Gravity}, Phys.
Rev. {\bf D54} (1996) 5049, [arXiv:hep.th/9511071].

\bibitem{DEGB2} T. Torii, H. Yajima and K.I. Maeda, {\it Dilatonic Black Holes with Gauss-Bonnet Term}, Phys. Rev. {\bf D55} (1997) 739,
[arXiv:gr-qc/9606034].

\bibitem{DEGB3} S. Alexeyev and M. Pomazanov, {\it Black hole solutions with dilatonic hair in higher curvature gravity}, Phys. Rev. {\bf D55} (1997)
2110, [arXiv:hep-th/9605106].

\bibitem{Malda} J. Maldacena, {\it The Large N Limit of Superconformal Field Theories and Supergravity},  
Adv. Theor. Math. Phys. {\bf 2} (1998) 231-252, [arXiv:hep-th/9711200].

\bibitem{TheKSS} P. Kovtun, D. T. Son and A. O. Starinets, \textit{%
Holography and hydrodynamics: diffusion on the stretched horizons}, JHEP
\textbf{0310} (2003) 064, [arXiv:hep-th/0309213].

\bibitem{TheKSS2} P. Kovtun, D. T. Son and A. O. Starinets, \textit{%
Viscocity in Strongly Interacting Quantum Field Theories from Black Hole
Physics,} Phys. Rev. Lett. \textbf{94} (2005) 111601, [arXiv:hep-th/0405231].


\bibitem{MartaK} Y. Kats and P. Petrov, \textit{Effect of curvature squared
corrections in AdS on the viscocity of the dual gauge theory}, JHEP \textbf{%
0901} (2009) 044, [arXiv:0712.0743].

\bibitem{RHIC} M. Brigante, H. Liu, R. Myers, S. Shenker and S. Yaida, 
\textit{The Viscosity Bound and Causality Violation}, [arXiv:0802.3318].

\bibitem{RHIC2} M. Brigante, H. Liu, R. Myers, S. Shenker and S. Yaida, 
\textit{Viscosity Bound Violation in Higher Derivative Gravity},
arXiv:0712.0805.

\bibitem{Dieguito} D. Hofman, \textit{Higher Derivative Gravity, Causality
and Positivity of Energy in a UV complete QFT}, [arXiv:0907.1625].

\bibitem{MyersApuraste} A. Buchel and R. Myers, \textit{Causality of
Holographic Hydrodynamics}, [arXiv:0906.2922].

\bibitem{Rut} R. Gregory, S. Kanno and J. Soda, {\it Holographic 
Superconductors with Higher Curvature Corrections}, [arXiv:0907.3203].

\bibitem{AdamsMaloney} A.~Adams, A.~Maloney, A.~Sinha and S.~E.~Vazquez, \textit{1/N Effects in Non-Relativistic 
Gauge-Gravity Duality}, JHEP {\bf 0903} (2009) 097, [arXiv:0812.0166].

\bibitem{DTSon} D.~T.~Son, \textit{Toward an AdS/cold atoms correspondence: a geometric realization of the 
Schroedinger symmetry}, Phys. Rev. {\bf D78} (2008) 046003, [arXiv:arXiv:0804.3972].

\bibitem{BalasubramanianMcGreevy} K.~Balasubramanian and J.~McGreevy, \textit{Gravity duals for non-relativistic CFTs}, 
Phys. Rev. Lett. {\bf 101} (2008) 061601, [arXiv:0804.4053].

\bibitem{Lanczos} C. Lanczos, \textit{A Remarkable property of the
Riemann-Christoffel tensor in four dimensions}, Ann. Math. \textbf{39}
(1938) 842.

\bibitem{CallanKlebanovPerry} C. Callan, I. Klebanov and M. Perry, \textit{%
String Theory Effective Actions}, Nucl. Phys. \textbf{B278} (1986) 78.

\bibitem{CandelasHorowitzStromingerWitten} P. Candelas, G. Horowitz, A.
Strominger and E. Witten, \textit{Vacuum Configurations for Superstrings},
Nucl. Phys. \textbf{B258} (1985) 46.

\bibitem{GrossSloan} D. Gross and J. Sloan, \textit{The Quartic Effective
Action for the Heterotic String}, Nucl. Phys. \textbf{B291} (1987) 41.

\bibitem{Zwiebach} B. Zwiebach, \textit{Curvature Squared Terms and String
Theories}, Phys. Lett. \textbf{B156} (1985) 315.

\bibitem{BoulwareDeser} D. Boulware and S. Deser, \textit{String Generated
Gravity Models}, Phys. Rev. Lett. \textbf{55} (1985) 2656.

\bibitem{CallanMyersPerry} C. Callan, R. Myers and M. Perry, \textit{Black
Holes in String Theory}, Nucl. Phys. \textbf{B311} (1989) 673.

\bibitem{Myers} R. Myers, \textit{Black Holes in Higher Curvature Gravity},
[arXiv:gr-qc/9811042].

\bibitem{Myers2} R. Myers, \textit{Superstring Gravity and Black Holes},
Nucl. Phys. \textbf{B289} (1987) 701.

\bibitem{Schiappa} F. Moura and R. Schiappa, \textit{Higher-Derivative
Corrected Black Holes: Perturbative Stability and Absorption Cross-Section
in Heterotic String Theory}, Class. Quant. Grav. \textbf{24} (2007) 361,
[arXiv:hep-th/0605001].

\bibitem{Otros} I. Bena and P. Kraus, \textit{R\symbol{94}2 Corrections to
Black Ring Entropy}, [arXiv:hep-th/0506015].

\bibitem{Fachoignorante} Y. Kats, L. Motl and M. Padi, \textit{Higher-order
corrections to mass-charge relation of extremal black holes}, JHEP \textbf{%
0712} (2007) 068, [arXiv:hep-th/0606100].

\bibitem{otros2} A. Sen, \textit{Entropy Function for Heterotic Black Holes}, JHEP \textbf{0603} (2006) 008, [arXiv:hep-th/0508042].

\bibitem{QWERT5} N. Dadhich, \textit{On the derivation of the gravitational
dynamics}, [arXiv:0802.3034].

\bibitem{MullerHoissen} F. M\"{u}ller-Hoissen, \textit{Spontaneous
Compactification With Quadratic And Cubic Curvature Terms}, Phys. Lett. 
\textbf{B163} (1985) 106.

\bibitem{TseytlinSeVolvioCubico} R. Metsaev and A. Tseytlin, {\it Curvature cubed terms in string theory
effective actions}, Phys. Lett. {\bf B185} (1987) 52.

\bibitem{T9907109} R. Troncoso and J. Zanelli, \textit{Higher Dimensional
Gravity, Propagating Torsion and AdS Gauge Invariance}, Class. Quant. Grav. 
\textbf{17} (2000) 4451.

\bibitem{Zanelli} J. Zanelli,\textit{\ Lecture notes on Chern-Simons
(super-)gravities}, 2nd edition, [arXiv:hep-th/0502193].

\bibitem{QWERT1} Deser, \textit{First-order Formalism and Odd-derivative
Actions}, Class. Quant. Grav. \textbf{23} (2006) 5773, [arXiv:gr-qc/0606006].

\bibitem{T0011097} R. Aros, R. Troncoso and J. Zanelli, \textit{Black holes
with topologically nontrivial AdS asymptotics}, Phys. Rev. \textbf{D63}
(2001) 084015, [arXiv:hep-th/0011097].

\bibitem{Wheeler} J. Wheeler, \textit{Symmetric Solutions to the
Gauss-Bonnet Extended Einstein Equations, }Nucl. Phys. \textbf{B268} (1986)
737.

\bibitem{Wheeler2} J. Wheeler, \textit{Symmetric Solutions To The Maximally
Gauss-Bonnet Extended Einstein Equations, }Nucl. Phys. \textbf{B273} (1986)
732.

\bibitem{Wiltshire} D. Wiltshire, \textit{Black Holes in String-Generated
Gravity Models}, Phys. Rev. \textbf{D38} (1988) 2445.

\bibitem{Wiltshire2} D. Wiltshire, \textit{Spherically Symmetric Solutions
Of Einstein-Maxwell Theory With A Gauss-Bonnet Term}, Phys. Lett. \textbf{%
B169} (1986) 36.

\bibitem{R41} Rong-Gen Cai, \textit{Gauss-Bonnet Black Holes in AdS Spaces},
Phys. Rev. \textbf{D65} (2002) 084014, [arXiv:hep-th/0109133].

\bibitem{R46} T. Jacobson, R. Myers, \textit{Entropy of Lovelock Black Holes}%
, Phys. Rev. Lett. \textbf{70} (1993) 3684, [arXiv:hep-th/9305016].

\bibitem{MyersBoundary} R. Myers, \textit{Higher Derivative Gravity, Surface
Terms And String Theory}, Phys. Rev. \textbf{D36} (1987) 392.

\bibitem{Davis} S. Davis, \textit{Generalised Israel Junction Conditions for
a Gauss-Bonnet Brane World}, Phys. Rev. \textbf{D67} (2003) 024030,
[arXiv:hep-th/0208205].

\bibitem{GravanisWillison2} E. Gravanis and S. Willison, \textit{Israel
conditions for the Gauss-Bonnet theory and the Friedmann equation on the
brane universe}, Phys. Lett. \textbf{B562} (2003) 118,
[arXiv:hep-th/0209076].

\bibitem{GravanisWillison} E. Gravanis and S. Willison, \textit{Intersecting
hypersurfaces in AdS and Lovelock gravity}, J. Math. Phys. \textbf{47}
(2006) 2503, [arXiv:hep-th/0412273].

\bibitem{GravanisWillison3} E. Gravanis and S. Willison, \textit{`Mass
without mass' from thin shells in Gauss-Bonnet gravity}, Phys. Rev. \textbf{%
D75} (2007) 084025, [arXiv:gr-qc/0701152].

\bibitem{GGGW} C. Garraffo, G. Giribet, E. Gravanis and S. Willison, \textit{%
Gravitational solitons and C\symbol{94}0 vacuum metrics in five-dimensional
Lovelock gravity}, J. Math. Phys. \textbf{49} (2008) 042503,
[arXiv:0711.2992].

\bibitem{QWERT2} Deser and A. Ryzhov, \textit{Curvature invariants of static
spherically symmetric geometries}, Class. Quant. Grav. \textbf{22} (2005)
3315, [arXiv:gr-qc/0505039].

\bibitem{R45} G. Allemandi, M. Francaviglia and M. Raiteri, \textit{Charges
and Energy in Chern-Simons Theories and Lovelock Gravity}, Class. Quant.
Grav. \textbf{20} (2003) 5103, [arXiv:gr-qc/0308019].

\bibitem{T0412046} P. Mora, R. Olea, R. Troncoso and J. Zanelli, \textit{%
Vacuum Energy in Odd-Dimensional AdS Gravity}, [arXiv:hep-th/0412046].

\bibitem{T0601081} P. Mora, R. Olea, R. Troncoso and J. Zanelli, \textit{%
Transgression forms and extensions of Chern-Simons gauge theories}, JHEP 
\textbf{0602} (2006) 067, [arXiv:hep-th/0601081].

\bibitem{Tcargas} R. Aros, M. Contreras, R. Olea, R. Troncoso and J.
Zanelli, \textit{Conserved Charges for Even Dimensional Asymptotically AdS
Gravity Theories}, Phys. Rev. \textbf{D62} (2000) 044002,
[arXiv:hep-th/9912045].

\bibitem{Tcargas2} R. Aros, M. Contreras, Rodrigo Olea, R. Troncoso and J.
Zanelli,\textit{\ Conserved charges for gravity with locally AdS asymptotics}%
, Phys. Rev. Lett. \textbf{84} (2000) 1647, [arXiv:gr-qc/9909015].

\bibitem{T0405267} P. Mora, R. Olea, R. Troncoso and J. Zanelli, \textit{%
Finite action principle for Chern-Simons AdS gravity}, JHEP \textbf{0406}
(2004) 036, [arXiv:hep-th/0405267].

\bibitem{R30} M. Dehghani, N. Bostani and A. Sheykhi, \textit{Counterterm
Method in Lovelock Theory and Horizonless Solutions in Dimensionally
Continued Gravity}, Phys. Rev. \textbf{D73} (2006) 104013, ,
[arXiv:hep-th/0603058].

\bibitem{R21} G. Kofinas and R. Olea, \textit{Universal regularization
prescription for Lovelock AdS gravity}, JHEP \textbf{0711} (2007) 069,
[arXiv:0708.0782].

\bibitem{R9} H. Maeda and M. Nozawa, \textit{Generalized Misner-Sharp
quasi-local mass in Einstein-Gauss-Bonnet gravity}, Phys. Rev. \textbf{D77}
(2008) 064031, [arXiv:0709.1199].

\bibitem{Olea} O. Miskovic and R. Olea, \textit{Counterterms in
Dimensionally Continued AdS Gravity}, [arXiv:0706.4460].

\bibitem{DeserMasa} S. Deser and B. Tekin, \textit{Gravitational Energy in
Quadratic Curvature Gravities}, Phys. Rev. Lett. \textbf{89} (2002) 101101,
[arXiv:hep-th/0205318].

\bibitem{H0303082} A. Padilla, \textit{Surface terms and the Gauss-Bonnet
Hamiltonian}, Class. Quant. Grav. \textbf{20} (2003) 3129,
[arXiv:gr-qc/0303082].

\bibitem{H0501044} N. Okuyama and J. Koga, \textit{Asymptotically anti de
Sitter spacetimes and conserved quantities in higher curvature gravitational
theories}, Phys. Rev. \textbf{D71} (2005) 084009, [arXiv:hep-th/0501044].

\bibitem{H0212292} S. Deser and B. Tekin, \textit{Energy in generic higher
curvature gravity theories}, Phys. Rev. \textbf{D67} (2003) 084009,
[hep-th/0212292].

\bibitem{H0310098} N. Deruelle, J. Katz and S. Ogushi, \textit{Conserved
charges in Einstein Gauss-Bonnet theory}, Class. Quant. Grav. \textbf{21}
(2004) 1971, [arXiv:gr-qc/0310098].

\bibitem{Deser} S. Deser and J. Franklin, \textit{Birkhoff for Lovelock Redux%
}, Class. Quant. Grav. \textbf{22} (2005) L103, [arXiv:gr-qc/0506014].

\bibitem{Zegres} R. Zegers, \textit{Birkhoff's theorem in Lovelock gravity},
J. Math. Phys. \textbf{46} (2005) 072502, [arXiv:gr-qc/0505016].

\bibitem{Cristo} C. Charmousis and J. Dufaux, \textit{General Gauss-Bonnet
brane cosmology}, Class. Quant. Grav. \textbf{19} (2002) 4671,
[arXiv:hep-th/0202107].

\bibitem{H0401192} V. Cardoso, O. Dias and J. Lemos , \textit{Nariai,
Bertotti-Robinson and anti-Nariai solutions in higher dimensions}, Phys.
Rev. \textbf{D70} (2004) 024002, [arXiv:hep-th/0401192].

\bibitem{Tangherlini} F. Tangherlini, \textit{Schwarzschild field in n
dimensions and the dimensionality of space problem}, Nuovo Cimento \textbf{27%
} (1963) 636.

\bibitem{BoulwareDeser2} D. Boulware and S. Deser, \textit{Effective Gravity
Theories with Dilatons}, Phys. Lett. \textbf{B175} (1986) 409.

\bibitem{Alan} A. Garbarz, G. Giribet and F.D. Mazzitelli, {\it Conformal invariance and apparent universality of 
semiclassical gravity}, Phys. Rev. {\bf D78} (2008) 084014, [arXiv:0807.4885]. 

\bibitem{R1} G. Dotti and R. Gleiser,\textit{\ Gravitational instability of
Einstein-Gauss-Bonnet black holes under tensor mode perturbations}, Class.
Quant. Grav. \textbf{22} (2005) L1, [arXiv:gr-qc/0409005].

\bibitem{R2} G. Dotti and R. Gleiser, \textit{Linear stability of
Einstein-Gauss-Bonnet static spacetimes. Part I: tensor perturbations},
Phys. Rev. \textbf{D72} (2005) 044018, [arXiv:gr-qc/0503117].\textit{\ }

\bibitem{R4} M. Beroiz, G. Dotti and R. Gleiser, \textit{Gravitational
instability of static spherically symmetric Einstein-Gauss-Bonnet black
holes in five and six dimensions}, [arXiv:hep-th/0703074].

\bibitem{DottiGleiser3} G. Dotti and R. Gleiser, \textit{Linear Stability of
Einstein-Gauss-Bonnet Spacetimes, Part II: Vector and Scalar Perturbations},
Phys. Rev. \textbf{D72} (2005) 124002, [arXiv:gr-qc/0510069].

\bibitem{R18} R. Konoplya and A. Zhidenko, \textit{(In)stability of
D-dimensional black holes in Gauss-Bonnet theory}, [arXiv:0802.0267].

\bibitem{H08043694} T. Hirayama, \textit{Negative modes of Schwarzschild
black hole in Einstein-Gauss-Bonnet Theory}, [arXiv:0804.3694].

\bibitem{R34} M. Dehghani and M. Shamirzaie, \textit{Thermodynamics of
Asymptotically Flat Charged Black Holes in Third Order Lovelock Gravity},
Phys. Rev. \textbf{D72} (2005) 124015, [arXiv:hep-th/0506227].

\bibitem{R28} M. Dehghani and N. Bostani, \textit{Spacetimes with
Longitudinal and Angular Magnetic Fields in Third Order Lovelock Gravity},
[arXiv:hep-th/0612103].

\bibitem{R53} R. Cai and N. Ohta, \textit{Black Holes in Pure Lovelock
Gravities}, Phys. Rev. \textbf{D74} (2006) 064001, [arXiv:hep-th/0604088].

\bibitem{CTZ} J. Crisostomo, R. Troncoso and J. Zanelli, \textit{Black Hole
Scan}, Phys. Rev. \textbf{D62} (2000) 084013, [arXiv:hep-th/0003271].

\bibitem{AFG} M. Aiello, R. Ferraro and G. Giribet, \textit{Exact Solutions
of Lovelock-Born-Infeld Black Holes}, Phys. Rev. \textbf{D70} (2004) 104014,
[arXiv:gr-qc/0408078].

\bibitem{R12} T. Torii and H. Maeda, \textit{Spacetime structure of static
solutions in Gauss-Bonnet gravity: charged case}, Phys. Rev. \textbf{D72}
(2005) 064007, [arXiv:hep-th/0504141].

\bibitem{R13} T. Torii and H. Maeda, \textit{Spacetime structure of static
solutions in Gauss-Bonnet gravity: neutral case}, Phys. Rev. \textbf{D71}
(2005) 124002, [arXiv:hep-th/0504127].

\bibitem{Hoffmann} B. Hoffmann, \textit{Gravitational and Electromagnetic
Mass in the Born-Infeld Electrodynamics}, Phys. Rev. \textbf{47} (1935) 877.

\bibitem{R6} G. Dotti, J. Oliva and R. Troncoso, \textit{Exact solutions for
the Einstein-Gauss-Bonnet theory in five dimensions: Black holes, wormholes
and spacetime horns}, [arXiv:0706.1830].

\bibitem{R3} G. Dotti and R. Gleiser, \textit{Obstructions on the horizon
geometry from string theory corrections to Einstein gravity}, Phys. Lett. 
\textbf{B627} (2005) 174, [arXiv:hep-th/0508118].

\bibitem{R36} M. Dehghani, \textit{Charged Rotating Black Branes in anti-de
Sitter Einstein-Gauss-Bonnet Gravity}, Phys. Rev. \textbf{D67} (2003)
064017, [arXiv:hep-th/0211191].

\bibitem{R35} M. Dehghani, \textit{Asymptotically (anti)-de Sitter solutions
in Gauss-Bonnet gravity with cosmological constant}, Phys. Rev. \textbf{D70}
(2004) 064019, [arXiv:hep-th/0405206].

\bibitem{R31} M. Dehghani and R. Mann, \textit{Thermodynamics of Rotating
Charged Black Branes in Third Order Lovelock Gravity and the Counterterm
Method}, Phys. Rev. \textbf{D73} (2006) 104003, [arXiv:hep-th/0602243].

\bibitem{R19} M. Dehghani, N. Alinejadi and S. Hendi, \textit{Topological
Black Holes in Lovelock-Born-Infeld Gravity}, [arXiv:0802.2637].

\bibitem{R8} M. Nozawa and H. Maeda, \textit{Dynamical black holes with
symmetry in Einstein-Gauss-Bonnet gravity}, Class. Quant. Grav. \textbf{25}
(2008) 055009, [arXiv:0710.2709].

\bibitem{QWERT12} S. Sarkar , S. Shankaranarayanan and L. Sriramkumar, 
\textit{Sub-leading contributions to the black hole entropy in the brick
wall approach}, [arXiv:0710.2013].

\bibitem{R54} S. Hendi and M. Dehghani, \textit{Taub-NUT Black Holes in
Third order Lovelock Gravity}, [arXiv:0802.1813].

\bibitem{R33} M. Dehghani and R. Mann, \textit{NUT-Charged Black Holes in
Gauss-Bonnet Gravity}, Phys. Rev. \textbf{D72} (2005) 124006,
[arXiv:hep-th/0510083].

\bibitem{R32} M. Dehghani and S. Hendi, \textit{Taub-NUT/Bolt Black Holes in
Gauss-Bonnet-Maxwell Gravity}, Phys. Rev. \textbf{D73} (2006) 084021,
[arXiv:hep-th/0602069].

\bibitem{Nueva} S. Habib Mazharimousavi and M. Halilsoy, \textit{5D black
hole solution in Einstein-Yang-Mills-Gauss-Bonnet theory}, Phys. Rev. D76 \
(2007) 087501, [arXiv:0801.1562].

\bibitem{Nueva2} S. Habib Mazharimousavi and M. Halilsoy, \textit{Black
Holes in Einstein-Maxwell-Yang-Mills Theory and their Gauss-Bonnet Extensions%
}, [arXiv:0801.2110].

\bibitem{QWERT8} A. Aliev, H. Cebeci and T. Dereli, \textit{Exact solutions
in five-dimensional axi-dilaton gravity with Euler-Poincare term}, Class.
Quant. Grav. \textbf{24} (2007) 3425, [arXiv:gr-qc/0703011].

\bibitem{R10} H. Maeda and N. Dadhich, \textit{Kaluza-Klein black hole with
negatively curved extra dimensions in string generated gravity models},
Phys. Rev. \textbf{D74} (2006) 021501, [arXiv:hep-th/0605031].

\bibitem{R15} A. Molina and N. Dadhich, \textit{On Kaluza-Klein spacetime in
Einstein-Gauss-Bonnet gravity}, [arXiv:0804.1194].

\bibitem{R16} N. Dadhich and H. Maeda, \textit{Origin of matter out of pure
curvature}, [arXiv:0705.2490].

\bibitem{R17} H. Maeda and N. Dadhich, \textit{Matter without matter: novel
Kaluza-Klein spacetime in Einstein-Gauss-Bonnet gravity}, Phys. Rev. \textbf{%
D75} (2007) 044007, [arXiv:hep-th/0611188].

\bibitem{R24} G. Giribet, J. Oliva and R. Troncoso, \textit{Simple
compactifications and Black p-branes in Gauss-Bonnet and Lovelock Theories},
JHEP \textbf{0605} (2006) 007, [arXiv:hep-th/0603177].

\bibitem{R25} D. Kastor and R. Mann, \textit{On black strings \& branes in
Lovelock gravity}, JHEP \textbf{0604} (2006) 048, [arXiv:hep-th/0603168].

\bibitem{H0412139} T. Kobayashi and T. Tanaka , \textit{Five-dimensional
black strings in Einstein-Gauss-Bonnet gravity}, Phys. Rev. \textbf{D71}
(2005) 084005, [arXiv:gr-qc/0412139].

\bibitem{H0509102} C. Sahabandu, P. Suranyi, C. Vaz and L. Wijewardhana, 
\textit{Thermodynamics of static black objects in D dimensional
Einstein-Gauss-Bonnet gravity with D-4 compact dimensions}, Phys. Rev. 
\textbf{D73} (2006) 044009, [arXiv:gr-qc/0509102].

\bibitem{HMyers} R. Myers and J. Simon, \textit{Black Hole Thermodynamics in
Lovelock Gravity}, Phys. Rev. \textbf{D38} (1988) 2434.

\bibitem{HWhitt} B. Whitt, \textit{Spherically Symmetric Solutions Of
General Second Order Gravity}, Phys. Rev. \textbf{D38} (1988) 3000.

\bibitem{MyersOtros1} T. Jacobson, G. Kang and R. Myers, \textit{Entropy
increase for black holes in higher curvature gravity}, Proceedings of the
7th Marcel Grossmann Meeting on General Relativity (1994) 937.

\bibitem{MyersOtros2} T. Jacobson, G. Kang and R. Myers, \textit{Increase of
black hole entropy in higher curvature gravity}, Phys. Rev. \textbf{D52}
(1985) 3518.

\bibitem{MyersOtros3} T. Jacobson, G. Kang and R. Myers, \textit{Black hole
entropy in higher curvature gravity}, [arXiv:gr-qc/9502009].

\bibitem{H9808067} R. Cai and K. Soh, \textit{Topological black holes in the
dimensionally continued gravity}, Phys. Rev. \textbf{D59} (1999) 044013,
[arXiv:gr-qc/9808067].

\bibitem{H0112045} M. Cvetic, S. Nojiri and S. Odintsov, \textit{Black hole
thermodynamics and negative entropy in de Sitter and anti-de Sitter
Einstein-Gauss-Bonnet gravity}, Nucl. Phys. \textbf{B628} (2002) 295,
[arXiv:hep-th/0112045].

\bibitem{H0202140} Y. Cho, I. Neupane, \textit{Anti-de Sitter black holes,
thermal phase transition and holography in higher curvature gravity}, Phys.
Rev. \textbf{D66} (2002) 024044, [arXiv:hep-th/0202140].

\bibitem{H0212092} I. Neupane, \textit{Black hole entropy in string
generated gravity models}, Phys. Rev. \textbf{D67} (2003) 061501,
[arXiv:hep-th/0212092].

\bibitem{H0302132} I. Neupane, \textit{Thermodynamic and gravitational
instability on hyperbolic spaces}, Phys. Rev. \textbf{D69} (2004) 084011,
[arXiv:hep-th/0302132].

\bibitem{R51} R. Cai, \textit{A Note on Thermodynamics of Black Holes in
Lovelock Gravity}, Phys. Lett. \textbf{B582} (2004) 237,
[arXiv:hep-th/0311240].

\bibitem{R29} M. Dehghani, G. H. Bordbar and M. Shamirzaie, \textit{%
Thermodynamics of Rotating Solutions in Gauss-Bonnet-Maxwell Gravity and the
Counterterm Method}, Phys. Rev. \textbf{D74} (2006) 064023,
[arXiv:hep-th/0607067].

\bibitem{R42} T. Clunan, S. Ross, D. Smith, \textit{On Gauss-Bonnet black
hole entropy}, Class. Quant. Grav. \textbf{21} (2004) 3447,
[arXiv:gr-qc/0402044].

\bibitem{QWERT6} S. Nojiri, S. Odintsov and S. Ogushi, \textit{%
Friedmann-Robertson-Walker brane cosmological equations from the
five-dimensional bulk (A)dS black hole}, Int. J. Mod. Phys. \textbf{A17}
(2002) 4809,[arXiv:hep-th 0205187].

\bibitem{QWERT7} J. Lidsey, S. Nojiri and S. Odintsov, \textit{\ Braneworld
Cosmology in (Anti)--de Sitter Einstein--Gauss--Bonnet--Maxwell Gravity},
JHEP \textbf{0206} (2002) 026, [arXiv:hep-th 0202198].

\bibitem{Sen} A. Sen, \textit{Black Hole Entropy Function and the Attractor
Mechanism in Higher Derivative Gravity}, JHEP \textbf{0509} (2005) 038,
[arXiv:hep-th/0506177].

\bibitem{R47} J. Morales and H. Samtleben, \textit{Entropy function and
attractors for AdS black holes}, JHEP \textbf{0610} (2006) 074,
[arXiv:hep-th/0608044].

\bibitem{R48} B. Chandrasekhar, S. Parvizi, A. Tavanfar and H. Yavartanoo, 
\textit{Non-Supersymmetric Attractors in R\symbol{94}2 Gravities}, JHEP 
\textbf{0608} (2006) 004, [arXiv:hep-th/0602022].

\bibitem{QWERT9} D. Astefanesei, H. Nastase, H. Yavartanoo, S. Yun, \textit{%
Moduli flow and non-supersymmetric AdS attractors}, JHEP \textbf{0804}
(2008) 074, [arXiv:0711.0036].

\bibitem{QWERT10} D. Astefanesei, K. Goldstein, R.P. Jena, A. Sen, S. P.
Trivedi, \textit{Rotating attractors}, JHEP \textbf{0610} (2006) 058,
[arXiv:hep-th/0606244].

\bibitem{R37} R. Cai, C. Chen, K. Maeda, N. Ohta and D. Pang, \textit{%
Entropy Function and Universality of Entropy-Area Relation for Small Black
Holes}, [arXiv:0712.4212].

\bibitem{Witten} E. Witten, \textit{(2+1)-Dimensional Gravity As An Exactly
Soluble System}, Nucl. Phys. \textbf{B311} (1988) 46.

\bibitem{QWERT3} S. Deser, R. Jackiw and S. Templeton, \textit{Topolgical
massive gauge theories}, Anals Phys. \textbf{140} (1982) 372, [Erratum-ibid.
185, 406.1988 APNYA, 281, 409 (1988) APNYA, 281-449.2000]-

\bibitem{QWERT4} S. Deser, R. Jackiw and S. Templeton, \textit{%
Three-dimensional massive gauge theories}, Phys. Rev. Lett. \textbf{48}
(1982)

\bibitem{T9601003} M. Ba\~{n}ados, R. Troncoso and J. Zanelli,\textit{\
Higher dimensional Chern-Simons supergravity}, Phys. Rev. \textbf{D54}
(1996) 2605, [arXiv:gr-qc/9601003].

\bibitem{BTZ} M. Ba\~{n}ados, C. Teitelboim and J. Zanelli, \textit{Geometry
of the (2+1) black hole}, Phys. Rev. \textbf{D48} (1993) 1506.
[gr-qc/9302012].

\bibitem{BTZbis} M. Ba\~{n}ados, C. Teitelboim and J. Zanelli, \textit{The
Black hole in three-dimensional space-time}, Phys. Rev. Lett. \textbf{69}
(1992) 1849, [hep-th/9204099].

\bibitem{BTZ4} M. Ba\~{n}ados, C. Teitelboim and J. Zanelli, \textit{%
Lovelock-Born-Infeld Theory of Gravity}, in J.J. Giambiagi Festschrift, Ed.
by H. Falomir et al. (1990).

\bibitem{BTZ2} M. Ba\~{n}ados, C. Teitelboim and J. Zanelli, \textit{%
Dimensionally continued black holes}, Phys. Rev. \textbf{D49} (1994) 975,
[arXiv:gr-qc/9307033].

\bibitem{BTZ3} M. Ba\~{n}ados, C. Teitelboim and J. Zanelli, \textit{Black
hole entropy and the dimensional continuation of the Gauss-Bonnet theorem},
Phys. Rev. Lett. \textbf{72} (1994) 957, [arXiv:gr-qc/9309026].

\bibitem{Giacomini} F. Canfora, A. Giacomini and R. Troncoso, \textit{Black
holes, parallelizable horizons and half-BPS states for the
Einstein-Gauss-Bonnet theory in five dimensions}, to appear in Phys. Rev. 
\textbf{D}, [arXiv:0707.1056].

\bibitem{Giacomini2} F. Canfora, A. Giacomini and S. Willison, \textit{Some
exact solutions with torsion in 5-D Einstein-Gauss-Bonnet gravity}, Phys.
Rev. \textbf{D76} (2007) 044021, [arXiv:0706.2891].

\bibitem{QWERT11} F. Canfora, \textit{Some solutions with torsion in
Chern-Simons gravity and observable effects}, [arXiv:0706.3538].

\bibitem{ROT2} A. Anabalon, N. Deruelle, Y. Morisawa, J. Oliva, M. Sasaki,
D. Tempo and R. Troncoso, \textit{Kerr-Schild ansatz in
Einstein-Gauss-Bonnet gravity: An exact vacuum solution in five dimensions},
to appear in Class. Quant. Grav. (2008), arXiv:0812.3194 [hep-th].

\bibitem{H08011021} Y. Brihaye and E. Radu, \textit{Five-dimensional
rotating black holes in Einstein-Gauss-Bonnet theory}, Phys. Lett. \textbf{%
B661} (2008) 167, [arXiv:0801.1021].

\bibitem{H08011021ZZZ} Hyeong-Chan Kim and Rong-Gen Cai, \textit{Slowly Rotating Charged Gauss-Bonnet Black holes in AdS Spaces}, 
Phys. Rev. {\bf D77} (2008) 024045, [arXiv:0711.0885].

\bibitem{HBhawal} B. Bhawal and S. Kar, \textit{Lorentzian wormholes in
Einstein-Gauss-Bonnet theory}, Phys. Rev. \textbf{D46} (1992) 2464.

\bibitem{R5} G. Dotti, J. Oliva and R. Troncoso, \textit{Static wormhole
solution for higher-dimensional gravity in vacuum}, Phys. Rev. \textbf{D75}
(2007) 024002, [arXiv:hep-th/0607062].

\bibitem{R7} H. Maeda and M. Nozawa, \textit{Static and symmetric wormholes
respecting energy conditions in Einstein-Gauss-Bonnet gravity},
[arXiv:0803.1704].

\bibitem{H07102041} M. Richarte and C. Simeone, \textit{Thin-shell wormholes
supported by ordinary matter in Einstein-Gauss- Bonnet gravity}, Phys. Rev. 
\textbf{D76} (2007) 087502, [arXiv:0710.2041].

\bibitem{COT} D. Correa, J. Oliva and R. Troncoso, \textit{Stability of
asymptotically AdS wormholes in vacuum against scalar field perturbations},
[arXiv:0805.1513].

\bibitem{Charmonius} C. Charmousis,\textit{\ Higher order gravity theories
and their black hole solutions}, [arXiv:0805.0568].

\bibitem{QWERT13} O. Miskovic and R. Olea, \textit{On boundary conditions in
three-dimensional AdS gravity}, Phys. Lett. \textbf{B640} (2006) 101,
[arXiv:hep-th/0603092].

\bibitem{R14} H. Maeda, \textit{Effects of Gauss-Bonnet term on the final
fate of gravitational collapse}, Class. Quant. Grav. \textbf{23} (2006)
2155, [arXiv:gr-qc/0504028].

\bibitem{R11} M. Nozawa and H. Maeda, \textit{Effects of Lovelock terms on
the final fate of gravitational collapse: analysis in dimensionally
continued gravity}, Class. Quant. Grav. \textbf{23} (2006) 1779,
[arXiv:gr-qc/0510070].

\bibitem{Gibbons} G. Gibbons, S. Hartnoll and A. Ishibashi, \textit{On the
Stability of Naked Singularities}, Prog. Theor. Phys. \textbf{113} (2005)
963-978, [arXiv:hep-th/0409307].

\bibitem{DottiNegative} G. Dotti and R. Gleiser , \textit{Instability of the
negative mass Schwarzschild naked singularity}, Class. Quant. Grav. \textbf{%
23} (2006) 5063-5078, [arXiv:gr-qc/0604021].

\bibitem{DottiBrasil} G. Dotti and M. Gleiser, privated communication, to be
presented at Seveth Alexander Friedmann Seminar on Gravitation and
Cosmology, Joao Pesoa, Brasil, 2008.

\bibitem{HorowitzMarolf} G. Horowitz and D. Marolf, \textit{Quantum probes
of space-time singularities}, Phys. Rev. \textbf{D52} (1995) 5670,
[arXiv:gr-qc/9504028 ].

\bibitem{Wald2} R. Wald, \textit{Dynamics In Nonglobally Hyperbolic, Static
Space-Times}, J. Math. Phys. \textbf{21} (1980) 2802.

\bibitem{IshibashiWald} A. Ishibashi and R. Wald, \textit{Dynamics in
Non-Globally-Hyperbolic Static Spacetimes II: General Analysis of
Prescriptions for Dynamics}, Class. Quant. Grav. \textbf{20} (2003)
3815-3826, [arXiv:gr-qc/0305012].

\bibitem{HosoyaIshibashi} A. Ishibashi and A. Hosoya, \textit{Who's Afraid
of Naked Singularities?}, Phys. Rev. D \textbf{60} (1999) 104028,
[arXiv:gr-qc/9907009].
\end{thebibliography}
\end{document}